\documentclass[reprint,amssymb,amsmath,aps,pra,longbibliography]{revtex4-1}
\usepackage{graphics}
\usepackage{bm}
\usepackage{epstopdf}
\usepackage{graphicx}
\usepackage{amsmath}
\usepackage{amssymb}
\usepackage{amsthm}
\usepackage{bm}
\usepackage{bbm}
\usepackage{graphics}

\begin{document}
\title{Ground state separability and criticality in 
interacting many-particle systems}
\author{Federico Petrovich$^1$, N.\ Canosa$^1$,  R.\ Rossignoli$^{1,2}$}
\affiliation{$^1$ Instituto de F\'{\i}sica de La Plata, CONICET, and Depto.\  de F\'{\i}sica, Facultad de Ciencias Exactas, Universidad Nacional de La Plata, C.C. 67, La Plata (1900), Argentina\\
$^{2}$ Comisi\'on de Investigaciones Cient\'{\i}ficas (CIC), La Plata (1900), Argentina}
\begin{abstract}
We analyze exact ground state (GS) separability in general $N$ particle  systems with two-site  couplings. General 
necessary and sufficient conditions for full separability, in the form of one and two-site eigenvalue equations, are first derived.  The formalism is then applied to a class of $SU(n)$-type interacting systems,  where each constituent has access to $n$ local levels, and where the total number parity of each level is preserved. Explicit factorization conditions for parity-breaking GS's are obtained,  which generalize those for $XYZ$ 
spin systems  and  correspond to a fundamental GS multilevel parity transition where the lowest $2^{n-1}$ energy levels cross. 
We also identify a multicritical factorization point with exceptional high degeneracy proportional to $N^{n-1}$, arising when the total occupation number of each level is preserved, in which {\it any} uniform product state is an exact GS. Critical entanglement properties (like full range 
pairwise entanglement) are shown to emerge  in the immediate vicinity of factorization.  Illustrative examples are provided. 
\end{abstract}
\maketitle

 \section{Introduction}
 \vspace*{-.45cm}
 
  The ground state (GS) of strongly interacting spin systems, while normally entangled \cite{ON.02,GVRK.03,A.08}, can exhibit the remarkable phenomenon of  factorization when a suitable magnetic field is applied 
\cite{Ku.82,MS.85,T.04,Am.06,Gi.08,RCM.08,RCM2.09,GG.09,Gi.09}. This means that for such field, the spin system admits a completely separable exact GS, i.e. a product of single spin states, despite the presence of nonnegligible couplings between the spins and the finite value of the applied field. Moreover, such product state is not necessarily trivial, in the sense that it may break fundamental symmetries of the Hamiltonian. 
In this case  factorization signals in finite systems a special critical point where two or more levels with definite symmetry cross and the GS becomes degenerate \cite{RCM.08,RCM2.09,GG.09,CRCR.17,CMR.20}, allowing for such  symmetry breaking exact eigenstates. The exact GS then typically undergoes in this case a transition between states with distinct symmetry as the factorization point is traversed, leading to visible effects in system observables 
\cite{RCM.08,RCM2.09,CRCR.17,CMR.20}. Furthermore, critical entanglement properties emerge in the immediate vicinity \cite{Am.06,RCM.08,RCM2.09,CRCR.17,CMR.20}, stemming ultimately from the product nature of the closely lying eigenstate. 

Most studies of GS factorization have so far been restricted  to interacting spin systems (see also \cite{RLA.10,CRC.10,Camp.13,CRC2.15}), where factorization conditions remain analytically manageable due to the small number of parameters required to specify an individual spin state. 
The main aim of this work is to investigate exact GS factorization  in more general interacting systems, i.e., beyond the standard $SU(2)$ spin scenario, where already the characterization of a single component state is more complex. With this goal, we first derive  the necessary and sufficient conditions for factorization in the form of eigenvalue equations, either for effective pair Hamiltonians or for the mean field (MF) Hamiltonian and residual couplings. 

We then apply  the formalism to  a general $N$-component  interacting system in which each constituent  has $n$ accessible local levels, such that the Hamiltonian can be expressed in terms of operators satisfying an $U(n)$ algebra. For $n=2$ it reduces to a general anisotropic $XYZ$ spin system  \cite{Baxter.71} 
in an applied transverse field \cite{Ku.82,T.04,RCM2.09,CRC2.15}, sharing with the latter the basic level number parity symmetry. For full range couplings it comprises  schematic $SU(n)$ models  employed in nuclear physics for describing collective excitations 
\cite{Mes.71,NPRC.85, RR2.87}, while for first neighbor couplings and special  choices of parameters it reduces to the $SU(n)$ Heisenberg model, also known as   Uimin-Lai-Sutherland (ULS) model \cite{Uimin.70,Lai.74,Sut.75}. 
The study of  interacting many body systems with 
global  $SU(n)$  symmetry  has aroused great  interest in recent years, becoming an active research topic  that links  the fields of  condensed matter and  atomic, molecular and optical  physics \cite{Sa.99,Ca.09,Gor.10,Ca.14,Lew.12,Taka.16}. 
Systems possessing high dimensional  symmetry  can  unveil exotic many body physics and are suitable for describing  a wide range of non-trivial  phenomena. The paradigmatic $SU(n)$ Heisenberg  model \cite{Uimin.70,Lai.74,Sut.75}, first employed in solid state physics in connection with the integer quantum Hall effect \cite{IA.85,IA.86}, played also  an important role  in identifying  unconventional magnetic states and phases 
\cite{HS.86, MA.89, RS.89, Re2.90, Gor.10, magne1.11, Du.15, Mi.18,YY.19}.
Interest on the subject has been stimulated by the unprecedented  advances in quantum control techniques, which  offer the possibility of realizing  strongly interacting many body systems with high symmetry in alkaline earth atomic gases in optical lattices \cite{Gor.10,Ca.09,Taka.16}. 
These  platforms have also received attention in relation with 
high precision atomic clocks \cite{Bloo.14} 
and  quantum computation \cite{Da.08}. 

The  general factorization  formalism  
is presented in section \ref{Formalism}, while its application to a general $SU(n)$-type model for $N$ components is described in \ref{III}. Explicit  equations for the existence of uniform parity-breaking factorized GS's are determined, and shown to correspond to a multilevel parity transition occurring for any size $N$ and coupling range, where the GS becomes $2^{n-1}$-fold degenerate (if $N\geq n-1$). A critical factorization point with exceptionally high degeneracy (which increases  with size $N$) is also identified in systems with full level number symmetry, where {\it any} uniform separable state is an exact GS. Entanglement properties in the vicinity of factorization  together with signatures of factorization in small systems are as well  discussed. Conclusions are drawn in \ref{IV}. Appendices discuss further details including the MF approximation in the  model, which admits an analytic solution in the uniform case for arbitrary $n$. 

\section{Formalism \label{Formalism}}
\subsection{General factorization conditions}
We consider a system described by a Hilbert space ${\cal H}=\bigotimes_{p=1}^{N}{\cal H}_{p}$,
such that it can be seen as a composite of $N$  subsystems with Hilbert spaces ${\cal H}_p$. 
 In this scenario we 
 assume  a general Hamiltonian containing
one-site terms $h_p$ plus two-site interactions $V_{pq}$: 
\begin{eqnarray}
H&=&\sum_{p}h_{p}+\frac{1}{2}\sum\limits _{p\neq q}V_{pq}\,,\label{H}\\
h_{p}&=&\sum\limits _{\mu}b_{\mu}^{p}o_{p}^{\mu}\,,\label{Hp}\;\;\;\;
V_{pq}=\sum\limits _{\mu,\nu}J_{\mu\nu}^{pq}o_{p}^{\mu}o_{q}^{\nu}\,,\label{Vpq}
\end{eqnarray}
where  $\left\{ o_{p}^{\mu}\right\} $ denotes a complete set of linearly independent
operators over ${\cal H}_p$ and    $J_{\mu\nu}^{pq}=J_{\nu\mu}^{qp}$ are  
the coupling strengths  of the interaction between sites
$p$ and $q$.  In particular, 
any spin array with two-spin
interactions in a general applied magnetic field  fits into this form. 
We use the notation $o_{p}^{\mu}\equiv\mathbbm{1}\otimes\cdots\otimes\mathbbm{1}\otimes o_{p}^{\mu}\otimes\mathbbm{1}\otimes\cdots\otimes\mathbbm{1}$
when operators are applied to global  states. 

We are here interested in  the  conditions which ensure that a completely separable  state 
\begin{equation}
\left|\Psi\right\rangle =\bigotimes_{p}\left|\psi_{p}\right\rangle =\left|\psi_{1},\ldots,\psi_{N}\right\rangle\,, \label{Psi}
\end{equation}
possibly breaking some fundamental  symmetry of $H$, 
is an \textit{exact} eigenstate of $H$: 
\begin{equation}
H\left|\Psi\right\rangle =E\left|\Psi\right\rangle .\label{HS}
\end{equation}
 When applied to $|\Psi\rangle$, $H$
can just connect it with itself and  with  superpositions of 
one-  and two-site ``excitations'',  
\begin{eqnarray}
\left|\Phi_{p}\right\rangle&=&\left|\psi_{1},\ldots,\phi_{p},\ldots\psi_{N}\right\rangle\label{phip}\,,\\
\left|\Phi_{pq}\right\rangle&=&\left|\psi_{1},
\ldots,\phi_{p},\ldots,\phi_{q}, \ldots,\psi_{N}\right\rangle\,, \label{phipq}
\end{eqnarray}
where 
$\left\langle \phi_{p}|\psi_{p}\right\rangle=\left\langle \phi_{q}|\psi_{q}\right\rangle=0$. Then  Eq. (\ref{HS}) implies the necessary and sufficient conditions 
\begin{eqnarray}
\left\langle \Phi_{p}\right|H\left|\Psi\right\rangle &=&0,\;\;\;p=1,\ldots,N\,,\label{1c}\\
\left\langle \Phi_{pq}\right|H\left|\Psi\right\rangle&=&0,\;\;\;1\leq p<q\leq N\,,\label{2c}
\end{eqnarray}
to be satisfied $\forall$  $\left|\phi_{p}\right\rangle$,
$\left|\phi_{q}\right\rangle$ 
orthogonal to $\left|\psi_{p}\right\rangle$,
$\left|\psi_{q}\right\rangle$ respectively. Since  
\begin{equation}
\left\langle \Phi_{p}\right|H\left|\Psi\right\rangle =\left\langle \phi_{p}\right|\tilde{h}_{p}\left|\psi_{p}\right\rangle\,,\;\;\tilde{h}_p=h_p+\sum_{q\neq p} v^{(q)}_p\,,\end{equation}
where $\tilde{h}_p$ is the local  MF  Hamiltonian at site $p$ and \begin{eqnarray}
 v^{(q)}_p&=&\langle\psi_q|V_{pq}|\psi_q\rangle=\sum_{\mu,\nu}\!J_{\mu\nu}^{pq}\left\langle o_{q}^{\nu}\right\rangle o_{p}^{\mu}\label{vpq}\,,\end{eqnarray} 
 the average  potential at $p$ due to the coupling with site $q$ 
($\langle o^\nu_q\rangle=\langle\psi_q|o^\nu_q|\psi_q\rangle$),  Eqs.\ (\ref{1c}) imply 
  $\langle \phi_p|\tilde{h}_p|\psi_p\rangle=0$ 
 $\forall$ $|\phi_p\rangle$ orthogonal to $|\psi_p\rangle$ and hence  the eigenvalue equations  \begin{equation}
\tilde{h}_{p}\left|\psi_{p}\right\rangle =\lambda_{p}\left|\psi_{p}\right\rangle ,\;\;p=1,\ldots,N.\label{HF}
\end{equation}
As expected, each local state $\left|\psi_{p}\right\rangle $ in $\left|\Psi\right\rangle $
should be an eigenstate of the local MF  Hamiltonian $\tilde{h}_p$ 
 determined by the same $\left|\Psi\right\rangle$, implying self-consistency. 

It is  now convenient to rewrite $H$ as 
\begin{equation}
    H=\sum_p \tilde{h}_p+\frac{1}{2}\sum_{p\neq q}\tilde{V}_{pq}\,,
    \label{Hhp}
\end{equation}
where  $\tilde{V}_{pq}=V_{pq}-v_p^{(q)}-v_q^{(p)}$ is a residual coupling  satisfying 
$\langle\Phi_p|\tilde{V}_{pq}|\Psi\rangle=\langle \Phi_q|\tilde{V}_{pq}|\Psi\rangle=0$. Then 
\begin{equation}
    \langle \Phi_{pq}|H|\Psi\rangle=
\langle \phi_{p},\phi_q|\tilde{V}_{pq}|\psi_p,\psi_q\rangle\,, 
\label{auxp}
\end{equation}
 and Eqs.\ \eqref{2c} together with previous property      
  imply that $|\Psi\rangle$ should  be an eigenstate  of all $\tilde{V}_{pq}$:  
\begin{equation}
\tilde{V}_{pq}
|\psi_{p},\psi_{q}\rangle =\lambda_{pq}|\psi_p,\psi_q\rangle\,,\;\;1\leq p<q\leq N\,,\\
\label{eqp}
\end{equation}
with  $\lambda_{pq}=\langle \tilde{V}_{pq}\rangle=-\langle V_{pq}\rangle$. As $\lambda_p=\langle h_p\rangle+\sum_{q\neq p}\langle V_{pq}\rangle$, the total energy verifies  
$E=\sum_p\lambda_p+\frac{1}{2}\sum_{p\neq q}\lambda_{pq}=\langle H\rangle$.
 
Therefore, we can state the following theorem:\\ 
{\it The product state $|\Psi\rangle$ 
is an exact eigenstate of the Hamiltonian (\ref{H}) iff $|\Psi\rangle$ is a simultaneous eigenstate of all one-site MF hamiltonians $\tilde{h}_p$ and  all residual couplings 
 $\tilde{V}_{pq}$.} 
 
Once Eqs.\ \eqref{HF} and \eqref{eqp} are fulfilled,  additional single site terms having $|\psi_p\rangle$ as GS  ($\Delta h_p|\psi_p\rangle=\Delta \lambda_p|\psi_p\rangle$) can be added to $H$ without affecting the product eigenstate.  They can be used 
to remove the eventual degeneracy  and bring down its energy ($E\rightarrow E+\sum_p \Delta\lambda_p$),   making it a nondegenerate GS for sufficiently large $\Delta \lambda_p<0$ $\forall\, p$.  

\subsection{Pair equations and the uniform case}
Eqs.\  \eqref{HF} and \eqref{eqp} imply that $H$ can be written as a sum of pair Hamiltonians $H_{pq}=H_{qp}$ ($p\neq q$) having the pair product state  $|\psi_p,\psi_q\rangle$ as eigenstate: 
\begin{eqnarray}
H&=&\frac{1}{2}\sum_{p\neq q}H_{pq}\,,\label{Hsp}\\
H_{pq}|\psi_p,\psi_q\rangle&=&E_{pq}|\psi_p,\psi_q\rangle\,\,,\;1\leq p<q\leq N\,.\label{Hpair}
\end{eqnarray}
For instance,  we can set 
$H_{pq}=r_{pq}(\tilde{h}_p +\tilde{h}_q)+\tilde{V}_{pq}$, 
with $r_{pq}=r_{qp}$ numbers satisfying  $\sum_{q}r_{pq}=1$ $\forall\,p$ (and $r_{pp}=0$)
in which case   
$E_{pq}=r_{pq}(\lambda_p+\lambda_q)+\lambda_{pq}$. 
The converse is trivially true:  Eqs.\ \eqref{Hsp}--\eqref{Hpair} imply Eq.\ \eqref{HS} for the 
state \eqref{Psi}, with 
\begin{equation}E=\frac{1}{2}\sum_{p\neq q}E_{pq}\label{Etpq}\,.\end{equation} 
Moreover, if $|\psi_p,\psi_q\rangle$ is a GS of $H_{pq}$ $\forall\,p\neq q$,  $|\Psi\rangle$  will clearly be {\it  a GS of $H$}, since it will minimize  
  {\it each average}  $\langle H_{pq}\rangle$ in \eqref{Hsp}, and hence the full average  $\langle H\rangle$. 

The pair Hamiltonians will have the general form 
 \begin{equation}
    H_{pq}=h_p^{(q)}+h_q^{(p)}+V_{pq}\,,\label{Hpair02}
\end{equation}
 with $\sum_{q\neq p}h^{(q)}_p=h_p$. Then, when multiplied by $\langle \psi_q|$, Eq.\
 \eqref{Hpair} leads to   $(h^{(q)}_p+v^{(q)}_p)|\psi_p\rangle=\lambda_p^{(q)}|\psi_p\rangle$, with $\lambda_p^{(q)}=E_{pq}-\langle h^{(p)}_q\rangle$, implying  Eq.\ \eqref{HF} when summed over $q$ (with $\lambda_p=\sum_q\lambda_p^{(q)}$) and 
 also Eq.\ \eqref{eqp} (with  $\lambda_{pq}=E_{pq}-\lambda^{(p)}_q-\lambda^{(q)}_p$). Eqs.\  \eqref{Hsp}--\eqref{Hpair} and \eqref{HF}--\eqref{eqp} are then equivalent. 
  
By expanding the local states $|\psi_p\rangle$ 
in an orthogonal basis,  
$|\psi_p\rangle=\sum_i f^p_i|i_p\rangle$  with  $f^p_i=\langle i_p|\psi_p\rangle$,  $\sum_i |f^p_i|^2=1$,  Eq.\ \eqref{Hpair} becomes, explicitly, 
 \begin{eqnarray}
 &&\sum_{j,l}[
  \delta_{kl}\langle i_p| h^{(q)}_p|j_p\rangle+
  \delta_{ij}\langle k_q| h^{(p)}_q|l_q\rangle+\langle i_pk_q|V_{pq}|j_pl_q\rangle]f^p_jf^q_l\nonumber
\\&&=E_{pq}f^p_if^q_k\label{eqx2},
\end{eqnarray}
  to be fulfilled $\forall$ $i,k$.  For  ${\rm dim}\,{\cal H}_{p(q)}=n_{p(q)}\geq 2$  and general couplings,  Eq.\ \eqref{eqx2} 
 imposes  $m=n_p n_q-1$ complex equations to be satisfied by product states $|\psi_{p},\psi_q\rangle$ 
having $l=n_p+n_q-2<m$ free complex parameters $f^p_i$, $f^q_j$, hence entailing restrictions on the feasible coupling strengths $J^{pq}_{\mu\nu}$ and ``fields'' $b^p_\mu$. Factorization will then take place at special ``points'' or ``curves'' in parameter space. In particular, If $H_{pq}$ is real in the previous pair product basis, one could always satisfy \eqref{eqx2} by adjusting the diagonal elements $\langle i_p k_q|V_{pq}|i_p k_q\rangle$. 

A simple realization of Eqs.\ \eqref{Hsp}--\eqref{Hpair} is the case of a uniform system where all local Hilbert spaces ${\cal H}_p$ and operators $o^\mu_p$ are identical, while  couplings between sites  are all proportional (or zero) such that $J^{pq}_{\mu\nu}=r_{pq}J_{\mu\nu}$ and  
\begin{eqnarray} V_{pq}&=&r_{pq}V\,,\;\;
V=\sum_{\mu,\nu}J_{\mu\nu}o^\mu\otimes o^\nu
\label{vpq2}\,,\\
h^{(q)}_p&=&r_{pq}h\,,\;\;
h=\sum_{\mu}b_\mu o^\mu\,,\label{rpq}\end{eqnarray}
in \eqref{Hpair02}, with  $V$ and $h$ {\it independent} of $p$ and $q$ (and $J_{\mu\nu}=J_{\nu\mu}$).  Here $r_{pq}=r_{qp}$ determines 
 the relative strength of the coupling between $p$ and $q$ and hence the range of the interaction.  Eqs.\ \eqref{vpq2}--\eqref{rpq} imply 
 \begin{eqnarray}
h_p&=&r_p h\,,\;\;r_p=\sum_{q\neq p} r_{pq}\,,\label{arp}\\
H_{pq}&=&r_{pq}(h\otimes \mathbbm{1}+\mathbbm{1}\otimes h+V)\,,
 \label{arpq}\end{eqnarray}
 such that all $H_{pq}$ become proportional. 

Then  a {\it uniform product eigenstate} with  $|\psi_p\rangle=|\psi\rangle$ $\forall p$ 
 may become feasible for special couplings, 
as {\it all}  pair equations  \eqref{Hpair} reduce in this case to  the {\it single} equation 
\begin{equation} (h\otimes \mathbbm{1}+\mathbbm{1}\otimes h+V)|\psi,\psi\rangle=E_2|\psi,\psi\rangle\,,\label{Ef}\end{equation}
after setting $E_{pq}=r_{pq} E_2$.   The total energy \eqref{Etpq} becomes 
\begin{equation}E=\frac{1}{2}E_2\sum_p r_p\,.\label{EEE}\end{equation}
Here $E_2$ represents a common pair  energy while $r_p$ a sort of coordination number for site $p$. In uniform  cyclic systems $r_p$ is constant  $\forall p$ and  $E=r_p\frac{N}{2}E_2$, while in open systems $r_p$ is typically  smaller at  the borders due to the smaller number of coupled neighbors, entailing edge corrections in $h_p=r_p h$. We will normalize the factors $r_{pq}$ such that $r_p=1$ for inner ``bulk''  sites (e.g.\ $r_{pq}=\frac{1}{2}\delta_{p,q\pm 1}$ for first neighbor couplings in a linear chain,  $r_{pq}=\frac{1}{N-1}$ for fully and equally connected systems). 

\subsection{Formulation for fermion and boson systems}
Previous equations admit a second quantized formulation for systems of fermions or bosons. For $N$ of such  particles at $N$ distinct (orthogonal) sites labelled by $p$, having each  $n_p={\rm dim}{\cal H}_p$ accessible local states labelled by $i$, we can define the corresponding creation and annihilation operators $c^\dag_{pi}$, $c_{pi}$ satisfying  \begin{equation}[c_{pi},c^\dag_{qj}]_{\pm}=
\delta_{pq}\delta_{ij}\,,\;[c^\dag_{pi},c^\dag_{qj}]_{\pm}=[c_{pi},c_{qj}]_{\pm}=0,\label{comm}\end{equation}
for fermions ($+$) or bosons ($-$) 
($[a,b]_{\pm}=ab\pm ba$). Setting  $o^\mu_p=g_p^{ij}=|i_p\rangle \langle j_p|$ and replacing it with  $c^\dag_{pi}c_{pj}$, we can express the equivalent of Hamiltonian \eqref{H} as 
\begin{equation}
    H=\sum_{p,i,j}b^{p}_{ij} c^\dag_{pi}c_{pj}+\frac{1}{2}\sum_{p\neq q}\sum_{i, j,k,l}\!\!J^{pq}_{ijkl}c^\dag_{pi}c^\dag_{qk}c_{ql}c_{pj}
    \label{Hn2}\,,\end{equation}
with $b^p_{ij}=\bar{b}^{p}_{ji}$, 
$J^{pq}_{ijkl}=J^{qp}_{klij}$ and $J^{pq}_{ijkl}=\bar{J}^{pq}_{jilk}$ for $H$ hermitian. 
It preserves the total occupancy at each site:  
\begin{equation} [H,N_p]=0\,,\;\;N_p=\sum_i c^\dag_{pi}c_{pi}\,,\label{s1}\end{equation}
(where $[a,b]=[a,b]_-$). We will consider  the single occupancy sector $N_p=1$ $\forall$ $p$, where the formulation in the previous form \eqref{H} is  equivalent. The  commutators 
\begin{equation}
    [c^\dag_{pi} c_{pj},c^\dag_{qk}c_{ql}]=\delta_{pq}(\delta_{jk} c^\dag_{pi}c_{pl}-\delta_{il}c^\dag_{pk}c_{pj})\label{Un}
\end{equation}
are  the same for  fermions and bosons  and are  identical to those satisfied by   $g_p^{ij}=|i_p\rangle\langle j_p|$   
($[g_p^{ij},g_q^{kl}]=\delta_{pq}(\delta_{jk}g_p^{il}-\delta_{il}g_p^{kj})$),  defining an  $U(n_p)$ algebra at each site.  

The product state \eqref{Psi} corresponds in the fermionic or bosonic scenario to an independent particle state 
\begin{equation}
    |\Psi\rangle=(\prod_p a^\dag_{p1})|0\rangle\,, \;\;\;a^\dag_{pj}=\sum_i U^p_{ji} c^\dag_{pi}\,,\label{psif}
\end{equation}
where $U^p_{ji}$ are the elements of a unitary matrix $U^p$ such that the same  relations \eqref{comm} are fulfilled by the new operators  $a^\dag_{pj}$, $a_{pi}$. 
   Then the one and two-site excitations \eqref{1c}--\eqref{2c} can be written as \begin{equation}|\Phi_p\rangle=a^\dag_{pi}a_{p1}|\Psi\rangle\,,\;\;|\Phi_{pq}\rangle=a^\dag_{pi}a^\dag_{qj}a_{q1}a_{p1}|\Psi\rangle\end{equation}
for $|\phi_p\rangle=a^\dag_{pi}|0\rangle$, $|\phi_q\rangle=a^\dag_{qj}|0\rangle$ and $i,j\geq 2$. 
Thus, we can employ expression 
\eqref{eqx2}   with $f^p_i=U^p_{1i}$ and  
\begin{equation}\langle i_p k_q|V_{pq}|j_pl_q\rangle
=J^{pq}_{ijkl}\,.\label{Jpq}\end{equation}

\section{Application to $n$-level models \label{III}}
\begin{figure}[t]
\centerline{\scalebox{.35
}{\includegraphics[trim={0cm 1cm 0cm 0.2cm},clip]
{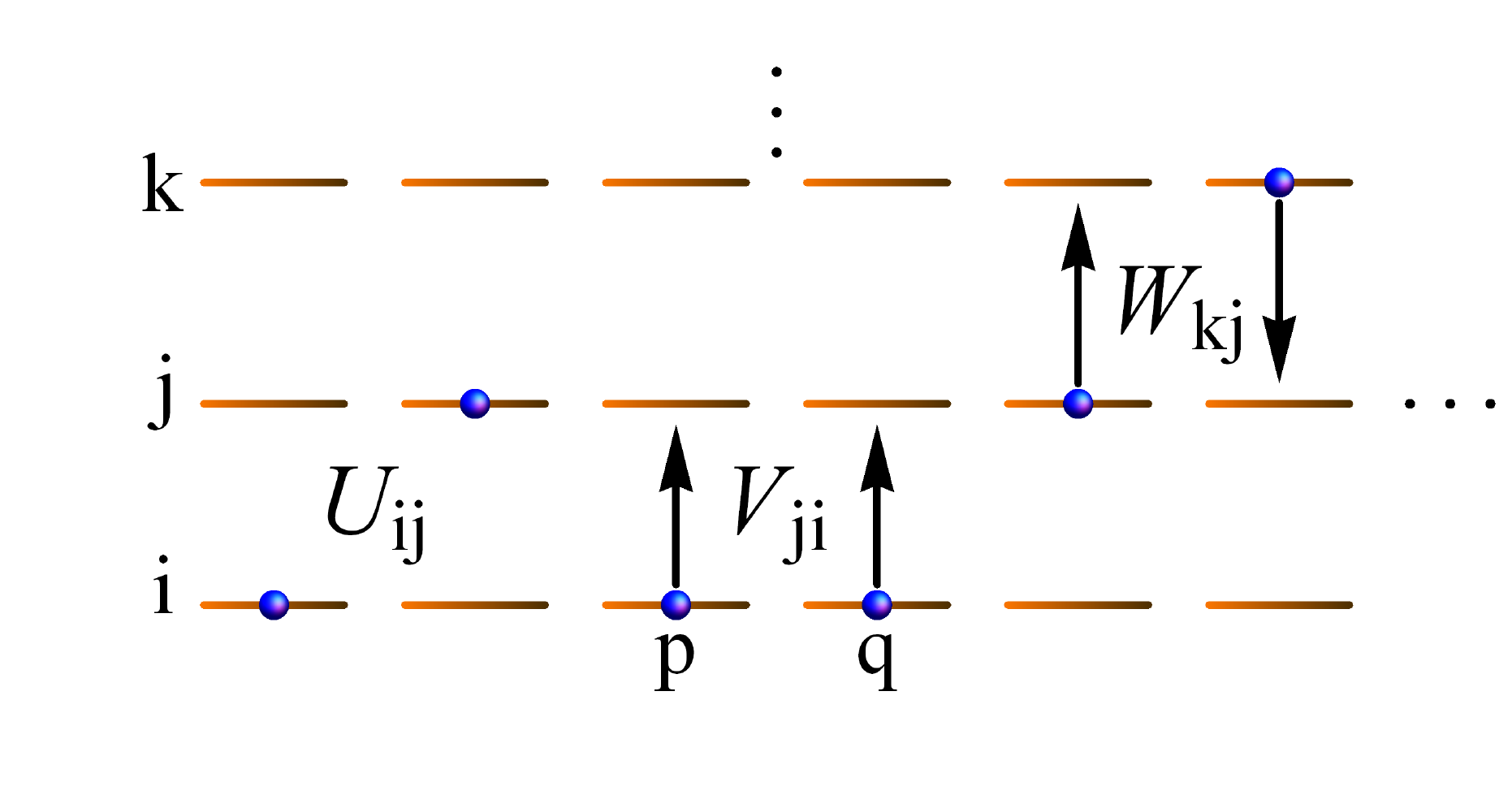}}}
\vspace*{-.25cm}
\caption{Schematic representation of the $U$, $V$ and $W$ couplings in the Hamiltonian \eqref{Hn3}.}  
\label{f1}
\end{figure}

We will now consider the problem of factorization in a general $n$-level  model with two-site interactions.    
It can be formulated as a system of  $N$ particles at $N$ distinct sites $p$, having each access to $n$ local  levels  with unperturbed energies $\epsilon_i^p$. 
The Hamiltonian reads  
\begin{equation}
\begin{aligned}
    H
    =&\sum_{i,p}\epsilon_i^p c^\dag_{pi}c_{pi}-
    \frac{1}{2}\sum_{p\neq q}r_{pq}\sum_{i,j}(U_{ij}c^\dag_{pi}c^\dag_{qj}c_{qj}c_{pi}+
    \\&+V_{ij}c^\dag_{pi}c^\dag_{qi}c_{qj}c_{pj}+W_{ij}
    c^\dag_{pi}c^\dag_{qj}c_{qi}c_{pj})\,,
      \end{aligned}\label{Hn3} \end{equation}
where $U_{ij}=U_{ji}$,  $V_{ij}=V_{ji}$ and  $W_{ij}=W_{ji}$ are real  coupling strengths 
and  $r_{pq}=r_{qp}$ determines the coupling range. The $V_{ij}$ terms promote two particles at sites $p,q$ from level $j$ to $i$, while the $W_{ij}$ terms interchange the occupancies of these levels at these sites (Fig.\ \ref{f1}). For $i=j$ both are identical to the $U_{ii}$ term so we set $V_{ii}=W_{ii}=0$ in what follows. The $U_{ij}$ terms just favor joint occupancy of levels $i,j$ at sites $p,q$. The operators $c^\dag_{pi}c_{pj}$ satisfy an $U(n)$ algebra at each site (Eq.\ \eqref{Un}). 

As discussed in App.\ \ref{ApA}, for full range couplings ($r_{pq}=\frac{1}{N-1}$ $\forall$ $p\neq q$)  the present model comprises the   fully connected  $SU(n)$ fermionic nuclear models  employed in \cite{Mes.71,NPRC.85,RR2.87}, which are an $n$-level generalization of the well-known two-level Lipkin model \cite{Lip,MDT.19}. Some $SU(n)$ spin models  and magnets \cite{magne1.11,magne2.16,magne3.20} also correspond to special cases of \eqref{Hn3}, with the $SU(n)$ invariant  Heisenberg coupling
\cite{Uimin.70,Lai.74,Sut.75,Ca.14,Du.15,Mi.18} recovered for  $V_{ij}=U_{ij}=0$ ($i\neq j$) and $W_{ij}=U_{ii}=J$. 
 In its  distinguishable formulation, \eqref{Hn3} is an $n$ level extension of the anisotropic $XYZ$ spin $1/2$ Hamiltonian in  an applied magnetic field 
  \cite{Ku.82,T.04,RCM2.09,CMR.20,Z.21}, recovered from \eqref{Hn3} for $n=2$. 
  Besides, for  $n=2s+1$ Eq.\ \eqref{Hn3} can  be formulated as a system of  spins $s$ with couplings depending on powers of the spin operators (see App.\ \eqref{ApA}). 
  
Since particles are moved in pairs between levels,  the  Hamiltonian  \eqref{Hn3} 
 has, for any value of the coupling strengths and range,  the  {\it number parity symmetries}
    \begin{align}[H,P_i]= &\; 0,\;\;i=1,\ldots,n\,,\label{PS}\\
    P_i= & \exp[-i\pi N_i]\,,\;\;N_i=\sum_p c^\dag_{pi}c_{pi}\,,
    \end{align}
    where $P_i$ is the parity of the total occupation $N_i$
 of level $i$. 
 Since $\prod_{i=1}^n P_i=e^{-i\pi N}$ is fixed,  just $n-1$ parities are independent. The exact eigenstates of $H$ will then have definite parities  when non-degenerate,  and can be characterized by their  $n-1$  values $\sigma_i=\pm 1$ for $i=2,\ldots,n$. 

  In the MF approximation,   
  which in the uniform attractive case  can be determined analytically (see App.\  \ref{ApB}) the  GS of \eqref{Hn3}  will typically exhibit a series of  transitions as  the coupling strengths increase from $0$, from the  unperturbed phase with all particles in the lowest  $i=1$ level,  to a final full parity-breaking phase where all $n$ levels are   occupied,  with  intermediate steps where  just $m<n$ levels are nonempty.  These transitions become  smoothed out in the actual entangled exact GS for finite $N$, which may instead exhibit number parity transitions (secs.\ \ref{IIIB} and \ref{IIIE}).  The parity-breaking MF GS becomes however {\it exact} at the factorization point,  discussed below. 

\subsection{Uniform factorized GS \label{IIIA}} 
 We now determine  the  conditions for which  the Hamiltonian  \eqref{Hn3} possesses  a uniform factorized GS 
     \begin{equation}
|\Psi\rangle=\prod_p a^\dag_{p1}|0\rangle\,,\;\;a^\dag_{p1}=\sum_{i}f_i c^\dag_{pi}\,,\label{psiuf}
\end{equation}
with $f_i$ $p$-independent and $\sum_i |f_i|^2=1$.  We set  $\epsilon_i^p=r_p\epsilon_i$ with $r_p=\sum_{q\neq p}r_{pq}$  according to \eqref{arp}, such that factorization is determined by the single Eq.\   
\eqref{Ef}. 

It is then seen that for $k=i$, Eq.\ \eqref{eqx2}  leads here to 
\begin{subequations}
\begin{equation} 
\sum_j[(2\epsilon_i-U_{ii})\delta_{ij}-V_{ij}]f_j^2=E_2 f_i^2\,,\label{eig1}
\end{equation}
for $i=1,\ldots,n$, which is a  {\it  standard eigenvalue equation} for the vector $\bm{f}^2$ of elements $f_i^2$ (i.e.,  for the ``squared wave function'') and matrix $M_{ij}=(2\epsilon_i-U_{ii})\delta_{ij}-V_{ij}$: 
\begin{equation}
M\bm{f}^2=E_2\bm{f}^2\,.
\label{eig20}
\end{equation}
\label{eigg}
\end{subequations}
It represents the $n\times n$ $ii$-$jj$ block in $\eqref{eqx2}$.  

On the other hand, for $k=j\neq i$, Eq.\ \eqref{eqx2} leads here to the $2\times 2$ $ij$-$ji$ block  
\begin{equation}
\!\!\begin{pmatrix}
\epsilon_i+\epsilon_j-U_{ij}&-W_{ij}\\
-W_{ij}&\epsilon_i+\epsilon_j-U_{ij}\end{pmatrix}\begin{pmatrix}f_i f_j\\f_j f_i\end{pmatrix}=E_2\begin{pmatrix}f_i f_j\\f_j f_i\end{pmatrix}\,.
\label{wij}
\end{equation}
Eq.\ \eqref{wij} entails,  for $f_i f_j\neq 0$, the constraint  \begin{equation} U_{ij}+W_{ij}=\epsilon_i+\epsilon_j-E_2\label{jij}\,.\end{equation}

Hence, given an arbitrary  single site spectrum 
$\epsilon_i$  and  couplings $V_{ij}$, $U_{ii}$,  
the factorized eigenstate and pair energy $E_2$ are first determined from the 
eigenvalue equation \eqref{eig20}. The couplings $W_{ij}$ or $U_{ij}$ 
for which such state becomes  an {\it exact} eigenstate are then obtained from   \eqref{jij}.  These conditions  are {\it independent} of coupling range $r_{pq}$ and system size $N$, implying that this  factorization will emerge for  {\it any  $N\geq 2$ and range $r_{pq}$} if  \eqref{jij} is  satisfied. The total energy is determined  by  $E_2$ through  Eq.\ \eqref{EEE}. 

For {\it GS factorization},  
 the {\it lowest} eigenvalue $E_2$  of \eqref{eigg} should be chosen. In this case, as the eigenvalues of the matrix in \eqref{wij} are $\epsilon_i+\epsilon_j-U_{ij}\mp W_{ij}$, i.e.\ $E_2$ and $E_2+2W_{ij}$ when \eqref{jij} is fulfilled, the uniform  factorized state  will be a GS of the full pair Hamiltonian (and hence of the full $H$) for any signs of the $V_{ij}$'s if  
  \begin{equation} 
  W_{ij}\geq 0\,\;\;\forall\,\,i\neq j
 \,,\label{ineq} 
 \end{equation}
 i.e.\ $E_2\leq \epsilon_i+\epsilon_j-U_{ij}$ $\forall\, i\neq j$. 
 Since the lowest eigenvalue of \eqref{eigg} satisfies $E_2\leq {\rm Min}_i[2\epsilon_i-U_{ii}]\leq 2\epsilon_i-U_{ii}$ $\forall\,i$,  a sufficient condition for the validity of \eqref{ineq} at  fixed $U_{ij}$ is 
 \begin{equation} 
U_{ij}\leq (U_{ii}+U_{jj})/2\,,
\label{ineq2} 
 \end{equation}
$\forall$ $i\neq j$. In particular, \eqref{ineq} will be always satisfied for the lowest eigenvalue $E_2$ if $U_{ij}=0$ $\forall$ $i,j$ and \eqref{jij} is fulfilled. 
The factorized GS obtained from \eqref{eigg} coincides, of course, with the MF GS for the couplings \eqref{jij}, lying {\it within} the full parity-breaking MF phase (see App.\ \ref{ApB}). 
 
For  $n=2$,  the factorization conditions 
\eqref{eigg}, \eqref{jij} reduce to those for the $XYZ$ spin Hamiltonian (see App.\ \ref{ApA}), leading to a factorizing field. And for $n=3$ it is still possible to satisfy \eqref{jij} by adjusting just the one-site energies $\epsilon_i$, for given values of $U_{ij}$ and $W_{ij}$:  
\begin{equation}
    \epsilon_i={\textstyle\frac{1}{2}}(T_{ij}+T_{ik}-T_{jk}+E_2)\,,\label{eid}
\end{equation}
where $T=U+W$ and $i\neq j\neq k$. In this case  a constant diagonal term  $\Delta U_{ii}=U_0$  remains to be added in \eqref{eigg} in order that  $E_{2}$ matches the original value.  

In the attractive case  $V_{ij}\geq 0$  $\forall$ $i,j$,  the eigenvector $\bm{f}^2$ of \eqref{eigg} associated to the lowest eigenvalue $E_2$ will have all components $f_i^2$ of the same sign (in order to yield the lowest eigenvalue) and hence all $f_i$ can be chosen as real.
Otherwise some of the $f_i^2$ can be negative, implying imaginary components $f_i$. 

In systems which can be divided into even and odd sites such that any site $p$ is coupled  ($r_{pq}\neq 0$) just to sites  $q$ of opposite parity (like first neighbor couplings in a linear chain or cubic lattice), the uniform factorized GS can be used to generate, through local unitaries,  {\it alternating} factorized GS's   for associated Hamiltonians. For instance, if 
$c^\dag_{pi}\rightarrow -c^{\dag}_{pi}$  for some level $i$ 
at odd sites $p$, then $V_{ij}\rightarrow -V_{ij}$, $W_{ij}\rightarrow -W_{ij}$ and 
 $|\Psi\rangle$ is changed into 
 an alternating product GS $|\Psi'\rangle$ with $f_i^p\rightarrow (-1)^p f_i$.    
 
    \subsection{Parity breaking and degeneracy at factorization\label{IIIB}}  
   Eqs.\ \eqref{eigg} just determine the squared coefficients $f_i^2$, leaving the sign of each  $f_i$ free. This degeneracy of the uniform factorized eigenstate \eqref{psiuf} reflects  its  breaking of   {\it all} number parity symmetries $P_i$ if  $f_i\neq 0$ $\forall \ i$:  Its expansion in the standard  ``product'' basis, 
\begin{equation}
    |\Psi\rangle=\sum_{i_1,\ldots,i_N}f_{i_1}\ldots f_{i_N}c^\dag_{1i_1}\ldots c^\dag_{Ni_n}|0\rangle\label{psiphf}
\end{equation}
clearly contains terms with all possible  parities $P_i$. As  
\begin{equation}
P_i c^\dag _{pi}P_i^\dag=-c^\dag _{pi}\,,
\label{Pci}
\end{equation}
$P_{i}|\Psi\rangle$ just changes 
the sign of $f_i$. Hence, if $|\Psi\rangle$ is an exact eigenstate, 
all $2^{n-1}$ parity transformed  states 
 \begin{equation}
     |\Psi_{i_1\ldots i_m}\rangle=P_{i_1}\ldots P_{i_m}|\Psi\rangle\,,
    \label{psip}
 \end{equation}
 obtained by changing the signs of $f_{i_1}\ldots f_{i_m}$ in \eqref{psiphf} with $m\leq n-1$,  are also exact eigenstates with the same energy due to \eqref{PS}. 
These parity breaking product eigenstates can then only arise at a point where levels with different parities {\it cross and become degenerate}.  Factorization  then signals 
 a {\it fundamental  parity level crossing} taking place for {\it any} size $N$ and range $r_{pq}$ whenever Eq.\ \eqref{jij} is fulfilled. 

If  $N \geq n-1$,  we thus obtain  from \eqref{psip}  $2^{n-1}$ nonorthogonal but linearly independent degenerate product eigenstates, implying   a  $D=2^{n-1}$ degeneracy at factorization,  which indicates the number of distinct parity levels exactly crossing  at this point.  
 
 On the other hand, for small systems with $N<n-1$,  the number $D$ of linearly independent states obtained with such sign changes in the $f_i$, and hence the degeneracy at factorization is smaller. We obtain in general  
 \begin{equation}D=\left\{\begin{array}{ccl}
 2^{n-1}&,&N\geq n-1\\\sum_{k=0}^N\binom{n-1}{k}&,&N\leq n-1\end{array}\right. , \label{deg}
 \end{equation}
 such that signs  are to  be changed in just $k\leq N$ levels. 
For a single pair ($N=2$),  $D=\binom{n}{2}+1$. 
 
 We have so far assumed that  the  matrix $M$ in \eqref{eigg} has  a non-degenerate GS, with a full rank eigenvector  $\bm{f}^2$.  If $f_i=0$ for some $i$, then factorization (and the ensuing degeneracy) becomes equivalent to that for $n\rightarrow n-1$.  And if the GS of $M$ is itself degenerate, the coefficients  $f_i^2$ will no longer be unique (after normalization). The GS of $H$ will then exhibit additional degeneracy, since a {\it continuous set} of factorized GS's becomes feasible.  We will consider  below  a special extreme case. 
 
 \subsection{The $W$-case: Number symmetry and exceptional degeneracy at factorization\label{Wsec}}
We  now consider the special case  where $V_{ij}=0$ $\forall$ $i\neq j$ in \eqref{Hn3}. 
For $n=2$ it  corresponds to the $XXZ$ model (see App.\ \ref{ApA}) which conserves the total $S_z$ and hence has eigenstates with definite  magnetization. Accordingly, for $V_{ij}=0$ Eq.\  \eqref{Hn3}  exhibits an additional symmetry: not only parity  but also the total occupation of each level  is conserved: 
\begin{equation} [H,N_i]=0\,,\;\;\;i=1,\ldots,n\,,\label{nii}\end{equation}
since the $U$ and $W$ couplings preserve all $N_i$'s. The exact eigenstates can then be characterized by the occupations $N_i$ of each level, existing  $\frac{N!}{N_1!\ldots N_n!}$ orthogonal states with the same set of occupations $(N_1,\ldots,.N_n)$.  

This higher  symmetry entails, first, 
a trivial factorization: the $n$ states with all particles in just one level, \begin{equation}
    |\Psi_i\rangle=\prod_p c^\dag_{pi}|0\rangle\,,\;\;\;i=1,\ldots,n\,,
\label{psin}
\end{equation} 
are clearly exact eigenstates: 
    $H|\Psi_i\rangle=E_i|\Psi_i\rangle$      with $E_i=(\epsilon_i-\frac{1}{2}U_{ii})\sum_p r_p$. 
    For $n=2$ they become the fully aligned spin states with maximum magnetization $|M|$. 

But in addition, {\it non-trivial symmetry-breaking} uniform factorized eigenstates of the form \eqref{psiuf} may also arise:  Eqs.\ \eqref{eigg}--\eqref{jij}  remain valid, but Eq.\   \eqref{eigg} becomes trivial, implying, for a full rank solution with $f_i\neq 0$ $\forall i$,  
\begin{eqnarray} U_{ii}&=&2\epsilon_i-E_2\,,\;\;i=1,\ldots,n\,,\label{wii}\\
W_{ij}+U_{ij}&=&\epsilon_i+\epsilon_j-E_2=
{\textstyle\frac{U_{ii}+U_{jj}}{2}}\label{jij2}\,.\end{eqnarray}
 Thus,  $f_i$ remains here {\it completely  arbitrary}: For vanishing $V_{ij}$ {\it any uniform  factorized state \eqref{psiuf} is an exact eigenstate with the same energy \eqref{EEE}} when \eqref{wii}--\eqref{jij2} are fulfilled, as the matrix $M$ becomes proportional to the identity.  And if $W_{ij}\geq 0$ $\forall$ $i\neq j$, i.e.\ if Eq.\ \eqref{ineq2} holds $\forall$ $i\neq j$, {\it they will be GS's} by the same previous arguments.  The ensuing GS energy \eqref{EEE} is then {\it independent} of the number $n$ of levels for a given fixed value of $E_2$. 
 
Such  {\it continuous} set of   factorized exact GS's  reflects  their breaking of {\it all} number symmetries \eqref{nii} when $0<f_i<1$ $\forall \ i$, as they lead to non-zero fluctuations  $\langle N_i^2\rangle-\langle N_i
\rangle^2=Nf_i(1-f_i)>0$. 
Moreover, since they  contain terms with all possible values $0\leq N_i\leq N$ when $f_i\neq 0$ $\forall$ $i$,  {\it all number  projected states with definite values} $N_i=n_i$  $\forall \ i$ 
derived from such  product state $|\Psi\rangle$, 
\begin{eqnarray}
|\Psi_{n_1\ldots n_n}\rangle\propto P_{n_1}\ldots P_{n_n}|\Psi\rangle\label{Pns}\,,
\end{eqnarray}
satisfying $N_i|\Psi_{n_1\ldots n_n}\rangle=n_i|\Psi_{n_1\ldots  n_n}\rangle$ with $\sum_{i=1}^n n_i=N$, will also be {\it exact eigenstates with the same energy} due to \eqref{nii}. 
Here $P_{n_i}=\frac{1}{2\pi}\int _0^{2\pi}e^{-\imath\phi(N_i-n_i)} d\phi$ are number projectors ($[P_{n_i},H]=0$ $\forall$ $i$). 

Remarkably, when normalized  these projected states  become  {\it independent} of the arbitrary coefficients $f_i$ determining the product state $|\Psi\rangle$,  since each term in their expansion \eqref{psiphf} 
will have exactly $n_i$ particles in level $i$ and hence all coefficients  become identical:  $f_{i1}\ldots f_{i_N}=\prod_{i=1}^n(f_i)^{n_i}=C_{n_1\ldots n_n}$. Therefore, the states \eqref{Pns}  become 
\begin{equation}
    |\Psi_{n_1\ldots n_n}\rangle=
|n_1\ldots n_n\rangle \,,\label{Pns2}
\end{equation}
where $|n_1\ldots n_n\rangle$ is the {\it fully symmetric} state having $N_i=n_i$ particles in each level $i$.  The total degeneracy at factorization is then  given by the number of  distinct projected states \eqref{Pns2},  which is just the  number of ways of distributing $N$ undistinguishable particles on $n$ levels: 
\begin{equation}
D=\binom{N+n-1}{n-1}\,,  \label{mw}
\end{equation}
with $D\approx \frac{N^{n-1}}{(n-1)!}$ for  $N\gg n$.  Then factorization arises at an {\it exceptional critical point where the  $D$ lowest levels with distinct values of the $N_i$'s  cross and become degenerate}. The ensuing  degeneracy {\it grows} with system size, in contrast with previous $N$-independent parity degeneracy.   

Since {\it any} uniform factorized state is an exact GS at the factorizing point,  the GS subspace is here clearly {\it invariant under  arbitrary $U(n)$  unitary transformations}  
\begin{equation}U=\exp[-i\sum_{i,j}T_{ij}\sum_p  c^\dag_{pi}c_{pj}]\,,\end{equation}
where $T$ is an arbitrary  hermitian matrix,  as $U$ transforms any product state \eqref{psiuf} into another uniform product state  and these states span the GS subspace: \begin{equation}|\Psi\rangle\rightarrow U|\Psi\rangle\; \Longrightarrow \;\bm{f}\rightarrow \exp[-iT]\bm{f}\,.\label{U}\end{equation}
 It corresponds to  $U=e^{-iT}
\!\otimes\ldots\otimes e^{-iT}$ in the distinguishable formulation. 

The question which now arises is whether the full $H$ also becomes $SU(n)$ invariant when the factorizing conditions \eqref{wii}--\eqref{jij2} are fulfilled. 
For $n=2$ this is indeed the case: as shown in App.\ \ref{ApA},  they  lead to a Heisenberg Hamiltonian  $H\propto -\sum_{p<q}r_{pq} \bm{s}_p\cdot\bm{s}_q$ plus constant terms, where $\bm{s}_p$ is the (dimensionless) spin operator at site $p$. Such $H$ is obviously invariant under arbitrary global rotations $e^{-i\phi\bm{k}\cdot\sum_p\bm{s}_p}$, with $\bm{k}$ an arbitrary unit vector, and admits any aligned product state $|\bm{k},\ldots,\bm{k}\rangle$, with $\langle\bm{k}|\bm{s}_p|\bm{k}\rangle=\frac{1}{2}\bm{k}$, as exact GS for arbitrary $\bm{k}$.  

However, for $n\geq 3$ 
{\it only the GS subspace remains invariant} in general, i.e., 
$[H,U]\neq 0$, with $[H,U]$ having just $D$ zero eigenvalues, corresponding to the GS subspace. Therefore, for $n\geq 3$ the general $SU(n)$ Heisenberg Hamiltonian
\cite{Uimin.70,Lai.74,Sut.75}
\begin{equation} H=-J\sum_{p<q}r_{pq}\sum_{i,j}c^\dag_{pi}c^\dag_{qj}c_{qi}c_{pj}\label{HSun}\end{equation}
is just a {\it particular case} of present 
factorizing Hamiltonian,  corresponding to $\epsilon_i=0$ $\forall$ $i$ and hence  
$U_{ii}=J=-E_2=W_{ij}$ $\forall$ $i\neq j$,  according to Eqs.\  \eqref{wii}--\eqref{jij2}.

 \subsection{Definite parity eigenstates and entanglement at the border of factorization}
We now examine the GS in the immediate vicinity of factorization. We consider first the $V\neq 0$ case. 
Since away from factorization the exact GS is normally non-degenerate for finite $N$, it will have definite parities $P_i$.  The same holds  for the other levels which meet at the factorization point. Therefore, their side-limits at factorization will be given by the parity projected states 
 \begin{equation}
     |\Psi_{\sigma_2\ldots\sigma_n}\rangle\propto(\mathbbm{1}+\sigma_2P_2)\ldots(\mathbbm{1}+\sigma_nP_n)|\Psi\rangle\,,\label{P}
 \end{equation}
where $\sigma_i=\pm 1$, satisfying  $P_i|\Psi_{\sigma_2\ldots\sigma_n}\rangle=\sigma_i |\Psi_{\sigma_2\ldots\sigma_n}\rangle$ $\forall i$.  This projection just selects from the expansion \eqref{psiphf} those terms  with the specified level parities. The GS will then exhibit a {\it  parity transition} as the factorization point is crossed \cite{RCM.08,RCM2.09,CMR.20} (when some Hamiltonian parameter is varied), having distinct parities $\sigma_i$  at each side. 

These projected states are {\it entangled}, i.e., they are no longer product  states. They exhibit critical entanglement properties since the product state $|\Psi\rangle$ from which they are derived is uniform and has lost all information about the range $r_{pq}$ of the coupling and the distance between sites.  Accordingly, 
the exact side-limits at factorization of GS entanglement entropies  will be {\it range-independent}. Moreover,   pairwise entanglement will be independent of the separation $|p-q|$ between sites, although it will remain small  in compliance with monogamy \cite{CKW.00,OTV.06}.  

These properties can be seen, for instance, in the reduced state of site $p$, $\rho_p={\rm Tr}_{p'\neq p}|\Psi_0\rangle\langle \Psi_0|$, 
of elements 
\begin{equation}
(\rho_p)_{ij}=\langle c^\dag_{pj}c_{pi}\rangle\,,
\end{equation}
and eigenvalues $\lambda_{pi}$. Its entropy \begin{equation}S_p=-{\rm Tr}\,\rho_p\log_2\rho_p=-\sum_{i=1}^n\lambda_{pi} \log_2 \lambda_{pi}\label{Sp}\end{equation}
is a measure of the (mode) entanglement between this site and  remaining sites. In the fermion case it is also a measure of {\it fermionic} entanglement \cite{GR.15,MDT.19}, in the sense of   indicating  the deviation of the state from an independent fermion state [Slater Determinant (SD)],  since it is the $p$-block  of the one-body density matrix $\rho^{(1)}$:  
\begin{equation}\rho^{(1)}_{pi,qj}=
\langle c^\dag_{qj}c_{pi}\rangle=\delta_{pq}\langle c^\dag_{pj}c_{pi}\rangle\,,
\end{equation}
whose  blocked structure  is due to  the fixed  fermion number $N_p$  at each site. Its entropy 
$S(\rho^{(1)})
=\sum_p S_p$ 
is a quantity which vanishes iff $|\Psi_0\rangle$ is a SD, i.e.\  $(\rho^{(1)})^2=\rho^{(1)}$ \cite{GR.15,DGR.18},  and is just $NS_p$ in the uniform case.  
In the factorized state $|\Psi\rangle$, $\langle c^\dag_{pj}c_{pi}\rangle=f^{p*}_i f^p_j$, implying  obviously $\rho_p^2=\rho_p$, i.e., $\lambda_{pi}=\delta_{i1}$, as directly seen in the MF basis ($\langle a^\dag_{pj}a_{pi}\rangle=\delta_{ij}\delta_{i1}$), and hence $S_p=0$.   

In contrast, in states $|\Psi_0\rangle$ with definite parity  all off-diagonal elements in the standard basis are cancelled by parity conservation ($[\rho_p,e^{i\pi c^\dag_{pi}c_{pi}}]=0$ $\forall$ $i$), implying 
\begin{equation}
  \langle c^\dag_{pj}c_{pi}\rangle=
  \delta_{ij}\langle c^\dag_{pi}c_{pi}\rangle\,.  \end{equation}
  Hence the eigenvalues of $\rho_p$ are just the average occupations 
   $\lambda_{pi}=\langle c^\dag_{pi}c_{pi}\rangle$  and  $S_p>0$ whenever      
 $\langle c^\dag_{pi}c_{pi}\rangle
\in (0,1)$.

In the projected states \eqref{P}, 
 these occupations depend on the parities $\sigma_2,\ldots,\sigma_n$.  For instance, for $n=3$  in the uniform case, we obtain, for $i=1,\ldots,3$, 
\begin{equation}\langle\Psi_{\sigma_2\sigma_3}|c^\dag_{pi}c_{pi}|\Psi_{\sigma_2\sigma_3}\rangle\
=|f_i|^2{\textstyle\frac{1+\sum_j(-1)^{\delta_{ij}}\sigma_j(1-2|f_j|^2)^{N-1}}{1+\sum_j 
\sigma_j(1-2|f_j|^2)^N}}\label{fp}
\end{equation}
where  $\sigma_1\sigma_2\sigma_3=(-1)^N$. Hence, for large $N$  
$\lambda_{pi}\approx |f_i|^2$ plus   corrections of order $(1-2|f_j|^2)^{N-1}$,   
which depend on the parities $\sigma_j$.  

For finite $N$ 
these corrections are, nonetheless, appreciable and their parity dependence  originates the splitting of the  degeneracy in the immediate vicinity of factorization (App.\ \ref{ApC}).  Moreover, the occupations \eqref{fp} determine the {\it exact side-limits} of the single-site entanglement entropy  \eqref{Sp} at factorization, which will then remain {\it finite} at this point and exhibit a {\it discontinuity}  due to the change in the GS parities $\sigma_i$. 
For large $N$ this discontinuity  becomes small,  as $\lambda_{pi}\approx |f_i|^2$ approaches the MF value at both sides, but the side-limits of $S_p$  remain {\it finite}. 

 On the other hand, the  entanglement between two sites $p\neq q$  is determined by their  reduced pair state $\rho_{pq}={\rm Tr}_{p'\neq p,q}|\Psi_0\rangle\langle\Psi_0|$, also a mixed state. For general $n$ it can be measured through the negativity \cite{VW.02,ZHSL.98,P.05}   
\begin{equation}
    {\cal N}_{pq}=\frac{1}{2}({\rm Tr}|\rho^{T_p}_{pq}|-1)\,,\label{neg}
\end{equation}
where $\rho^{T_p}_{pq}$ is the partial transpose of $\rho_{pq}$. Eq.\ \eqref{neg} is just minus the sum of the negative eigenvalues of $\rho^{T_p}_{pq}$, with ${\cal N}_{pq}>0$ ensuring entanglement of $\rho_{pq}$ 
according to  Peres criterion \cite{P.96}. 
The side-limits at factorization of the exact GS  negativities  will be determined by the projected states \eqref{P}, and  will  be {\it non-zero} for finite $N$, and  hence {\it independent} of the separation between  sites and the coupling range for a uniform $|\Psi\rangle$, undergoing there a discontinuity due to the transition in the GS parities. 

While visible in small systems (see  sec.\ \ref{IIIE}),  the common value of ${\cal N}_{pq}$ at factorization  decreases as $N$ increases, in agreement with monogamy: The projected states \eqref{P} involve a sum over $2^{n-1}$ product states 
$\sigma_{i_1}P_{i_1}\ldots \sigma_{i_m}P_{i_m}|\Psi\rangle$ having the signs of $f_i$ changed at levels $i_1,\ldots,i_m$, which for sufficiently large $N$ become approximately orthogonal (e.g.\ for $n=3$ their overlaps are proportional to terms  $(1-2|f_j|^2)^N$, as seen in \eqref{fp}, which decrease rapidly with $N$ if $|f_j|\neq 0$ or $1$). Neglecting these overlaps, the  two-site reduced states $\rho_{pq}$ derived from   \eqref{P} become essentially a  convex mixture of $2^{n-1}$ product states $\rho_p\otimes \rho_q$, and are then {\it separable} \cite{P.96}, implying ${\cal N}_{pq}\approx 0$ $\forall$ $p,q$. Thus, for large systems pairwise entanglement vanishes at factorization, though it will still show long range in its vicinity \cite{Am.06,RCM2.09,CMR.20}.

 We remark, however, that the exact GS side-limits at factorization of other entanglement measures do  remain finite for large $N$, as was seen for  the single site entropy \eqref{Sp}. In fact,  previous argument  entails  that the reduced state $\rho_M\equiv \rho_{p_1\ldots p_M}$  of $M<N$ sites derived from  \eqref{P} will be mixed with rank $2^{n-1}$ (for $M\geq n-1$),  such that its entropy, measuring their entanglement with the rest of the system, will also have  non-zero side-limits 
  for any $N$.  They will be bounded, however, by this rank:  
  \begin{equation}S(\rho_M)=-{\rm Tr}\,\rho_M\log_2\rho_M\leq n-1\,,\label{SMbnd}
  \end{equation}
  at the border of factorization. This bound at this point is then another signature of factorization in these systems.

 Similar considerations hold for the $V=0$ case. The level number projected states \eqref{Pns}--\eqref{Pns2} represent the exact side-limits at factorization of the $D$ crossing states. Except for the states \eqref{psin} with just one level occupied, all remaining states are entangled and lead again to critical entanglement properties (independence of coupling range and separation) due to their fully symmetric nature. In particular, they lead again to single site reduced states $\rho_p$ diagonal in the standard basis,  
\begin{equation}\langle n_1\ldots n_n| c^\dag_{pi}c_{p_j}|n_1\ldots n_n\rangle=\delta_{ij}n_i/N\,,\end{equation}
implying $\lambda_{pi}=n_i/N$ and hence a single-site entropy $S(\rho_p)>0$ if $1\leq n_i\leq N-1$ at least for some $i$.

\subsection{Factorization signatures in small systems\label{IIIE}}

\begin{figure}[t]
\centerline{\scalebox{.6}{\includegraphics{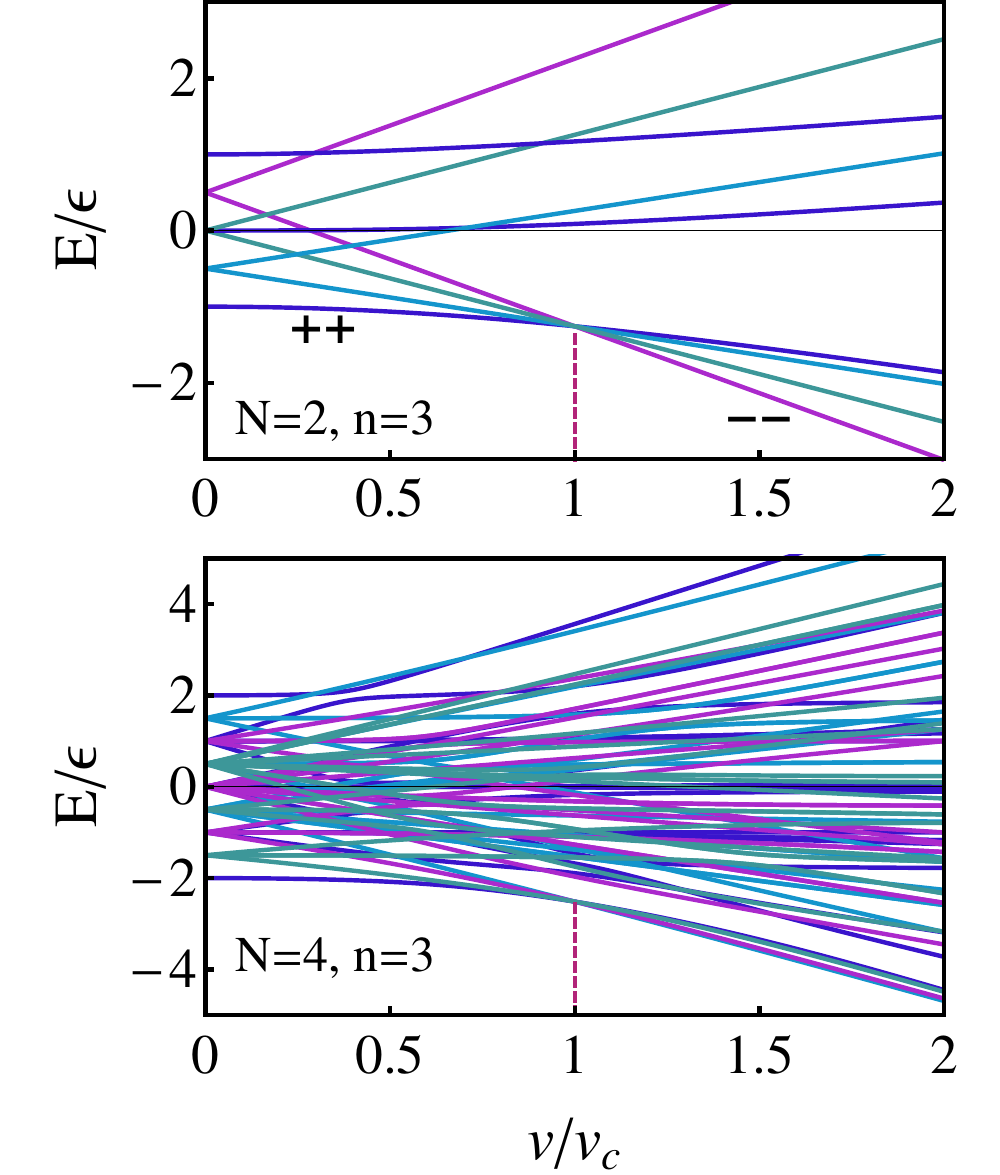}}}
\caption{The exact spectrum of Hamiltonian \eqref{Hn3} for a single pair ($N=2$, top)  and for   $N=4$ sites (bottom), with first neighbor couplings and   $n=3$ levels at each site, as a function of the scaled coupling strength  $v/v_c$ (see text).  In both cases factorization takes place at the same value $v=v_c$,   where Eqs.\ \eqref{eigg}-\eqref{jij} are fulfilled and the four levels with distinct parities forming the GS band cross.}
\label{f2}
\end{figure}

\begin{figure}[t]
\centerline{\scalebox{.4}{\includegraphics[trim={0cm 0cm 0 0cm},clip]{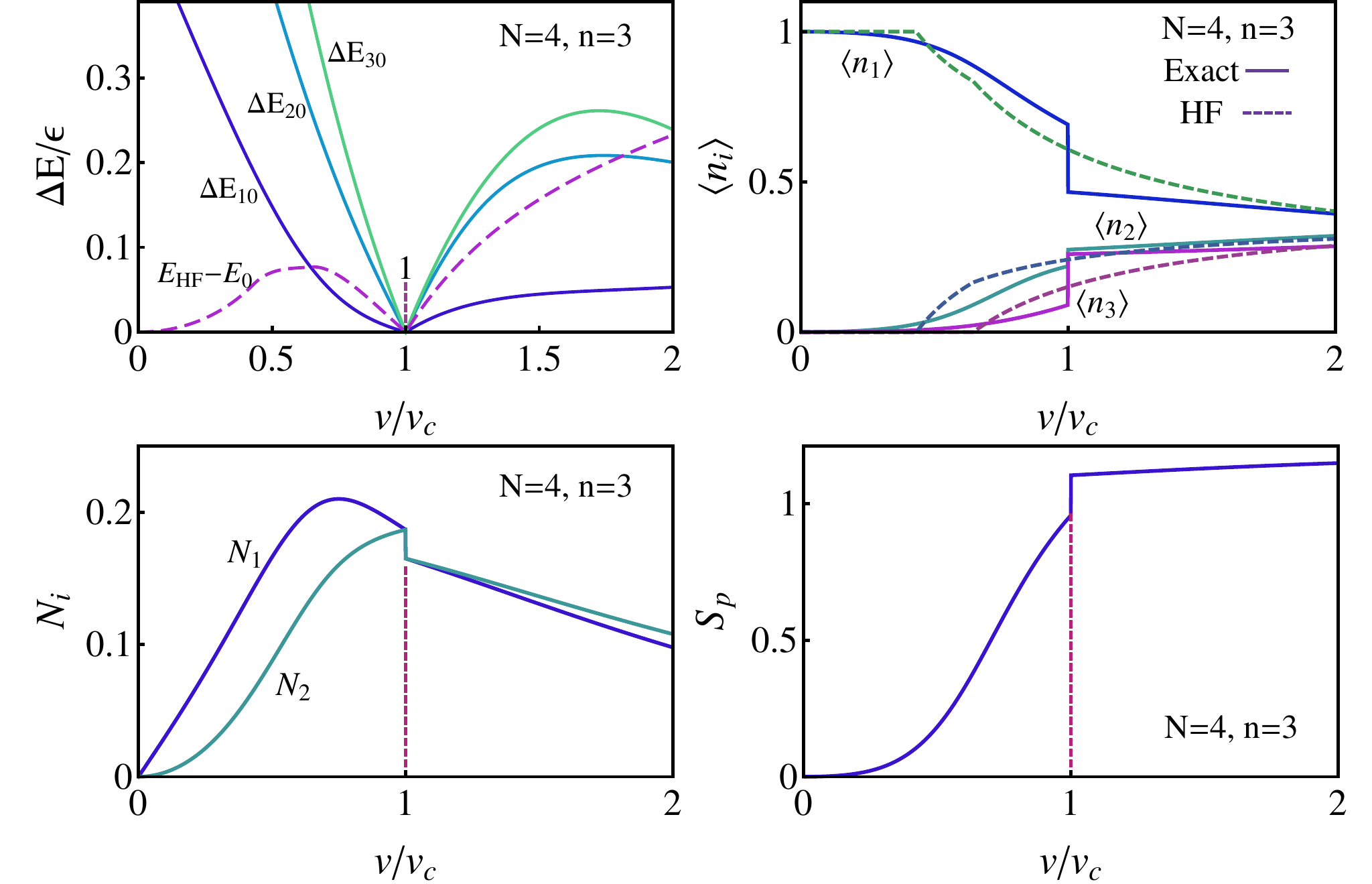}}}
\caption{Results for the $N=4$ chain  of  Fig.\ \ref{f2}. 
Top left: The first three  exact excitation energies $\Delta E_{i0}=E_i-E_0$  and  the difference $E_{\rm HF}-E_0$ with the Hartree-Fock (HF) GS energy. All vanish at the factorization point $v=v_c$ (1). 
Top right: Exact (solid lines) and  HF (dotted lines) values of the GS average occupations  $\langle n_i\rangle=\langle c^\dag_{pi}c_{pi}\rangle$ of the three levels. 
The exact values represent the eigenvalues of the single site reduced density matrix and exhibit a discontinuity  at $v=v_c$.  
  Bottom: The exact one-site entanglement entropy \eqref{Sp} (right), which shows a stepwise increase at factorization, and the exact negativities between first (${\cal N}_1={\cal N}_{p,p+1}$) and second (${\cal N}_2$) neighbors (left), measuring pairwise entanglement. Both reach the same side-limits at factorization, exhibiting there a stepwise decrease.}  \label{f3}
\end{figure}

\begin{figure}[t]
\centerline{\scalebox{.4}{\includegraphics[trim={0cm 0 0 0},clip]{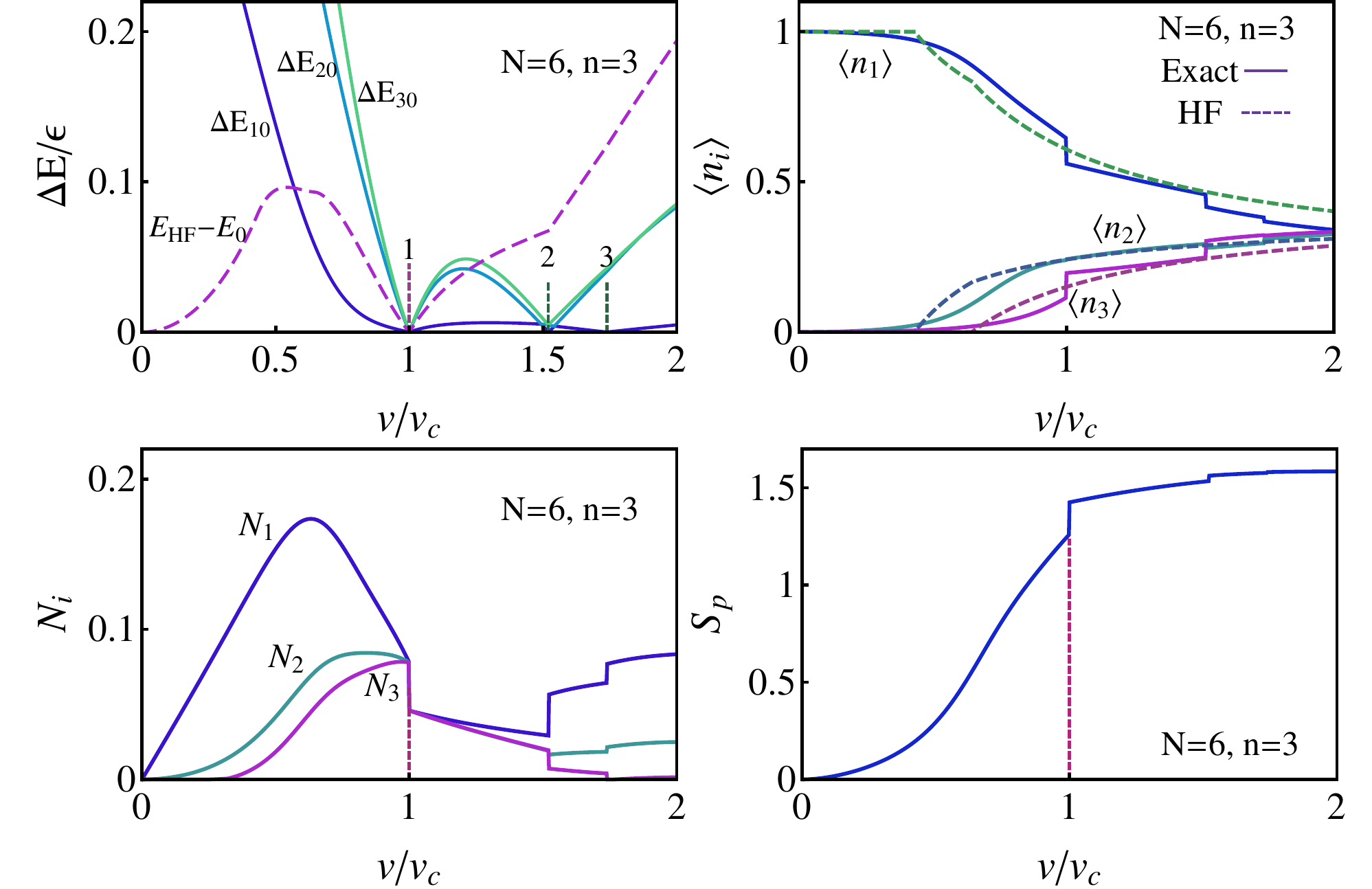}}}
\caption{Results for an $N=6$ chain with $n=3$ levels at each site. 
Details are similar to those of Fig.\ \ref{f3}.  Top left: The first three  excitation energies $\Delta E_{i0}$  together with 
 $E_{\rm HF}-E_0$. 
 Points $2,3$ indicate other  GS  parity transitions.
 Top right: Exact and  HF  average occupations $\langle n_i\rangle$. 
 Bottom:
The one-site entanglement entropy \eqref{Sp} (right)   
and the exact  negativities between first, second and third (${\cal N}_3$) neighbors (left).    
All ${\cal N}_i$ reach  the same side-limits just at factorization ($v=v_c$).}
\label{f4}
\end{figure}

\begin{figure}[t]
\centerline{\scalebox{.45}{\includegraphics[trim={0cm 0 0 0},clip]
{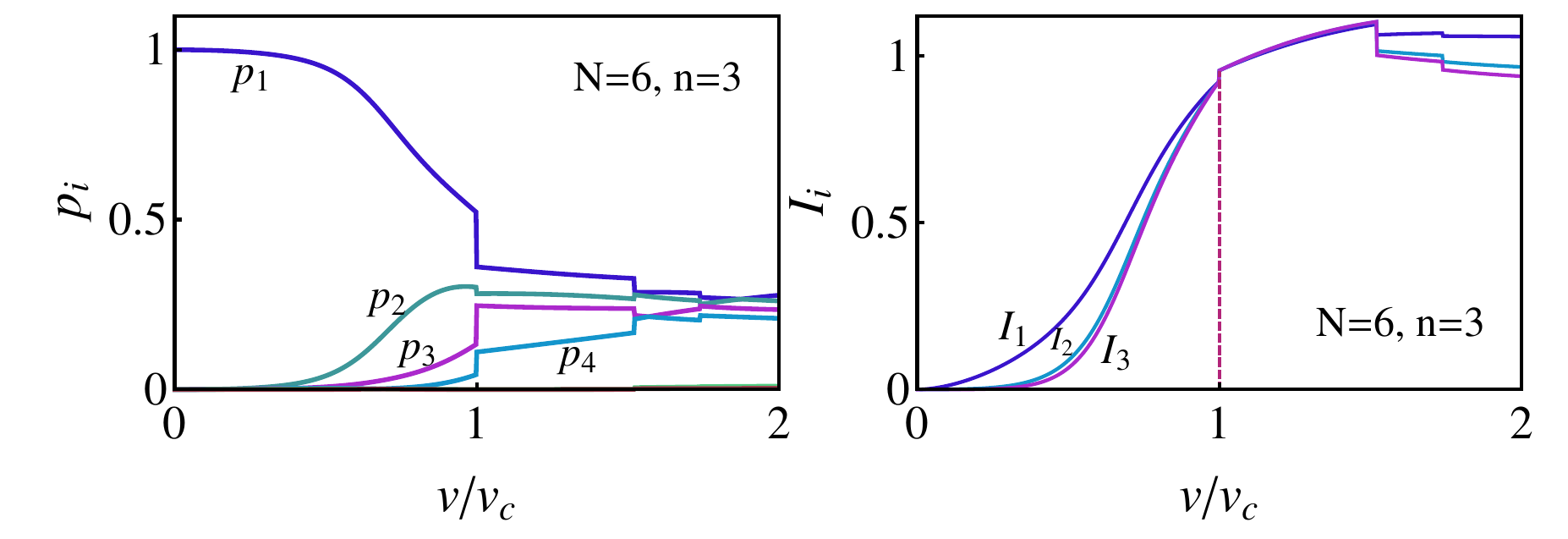}}}
\caption{The exact eigenvalues of the  two-site reduced density matrix for  first neighbors  (left) and the  mutual information $I(\rho_{pq})$ for first  ($I_1$), second  ($I_2$) and third  ($I_3$) neighbors (right), in the chain of Fig.\ \ref{f4}.
All $I_i$ exactly merge at the side-limits of the factorizing point $v=v_c$.}
\label{f5}
\end{figure}

We discuss here  typical illustrative results in small $n$-level systems.   We examine first the case with both $V$ and $W$ couplings  of sections  \ref{IIIA}-\ref{IIIB}. We consider a uniform single site spectrum $\epsilon_i=\frac{\epsilon}{2}(i-\frac{n+1}{2})$ for $i=1,\ldots,n$,  and  couplings  $U_{ij}=0$, $V_{ij}=v$  and $W_{ij}=(v/v_c)(\epsilon_i+\epsilon_j-E_{2c})$,  chosen such that GS factorization is reached at $v=v_c$, according to Eq.\ \eqref{jij} ($E_{2c}$ is the  pair energy obtained from \eqref{eigg} at $v=v_c$). For $n=2$ these parameters lead to  an anisotropic $XY$ Heisenberg coupling in a uniform field (Eq.\ \eqref{H2} with $J_z=0$), while for general $n$ it is an extension of the  $n$-level model used in \cite{NPRC.85,RR2.87}. Figs.\ \ref{f2}--\ref{f5} show results for the $n=3$-level case  with $v_c=\frac{2}{5}\epsilon$ (for which $E_{2c}\approx-1.26\epsilon$). 

We first depict in Fig.\ \ref{f2} the spectrum of $H$ for a single pair 
($N=2$, $r_{12}=1$) and for a cyclic four-particle chain with first-neighbor couplings ($N=4$,  $r_{pq}=\frac{1}{2}\delta_{q,p\pm 1}$),  as a function of $v/v_c$.  In  both cases there is a GS band of $2^{n-1}=4$ states which cross exactly at the  factorization point $v=v_c$, where  
a GS number parity transition takes place: The GS changes from the $(\sigma_1,\sigma_2)=(+,+)$ state for $v<v_c$, to the $(\sigma_1,\sigma_2)=(-,-)$ state for $v>v_c$. These states form the border of the GS band, the remaining crossing levels $(\sigma_1,\sigma_2)=(\pm,\mp)$ lying  in between.

Further results for a ring of $N=4$ particles are shown in Fig.\ \ref{f3}. It is verified that the first three exact excitation energies,  
 together with the difference with the mean field  (HF, see App.\ \ref{ApB}) GS energy, exactly vanish just at   $v=v_c$ (top left), confirming factorization. The exact average occupations   $\langle n_i\rangle$ of each level are shown in the top right panel (solid lines). As $v$ increases the two upper levels start to be populated,  with all exact occupations undergoing a  step-like discontinuity  at the factorizing point, reflecting the associated GS parity transition. The side-limits at this point  coincide  with those determined by the projected states \eqref{P} through Eq.\ \eqref{fp}. Present factorization can then be detected and verified through the magnitude of these occupation jumps.

 HF results reproduce qualitatively the general trend but miss the jump  at factorization: Though  exact  at this point,  the HF  GS corresponds to a superposition of the crossing definite parity exact eigenstates. It  exhibits  instead transitions at $v/v_c \approx 0.44$ and $0.65$ ($\forall\,N$), where the second and third level respectively  start to be populated in the approach (see App.\ \ref{ApB}) and parity symmetry becomes broken. Thus, factorization lies {\it within} the full parity-breaking HF phase (and not at a HF transition).  
 
Entanglement properties are depicted in the lower panels. The  exact single site entanglement entropy \eqref{Sp} (bottom right)  increases monotonously as $v/v_c$ increases, and  displays a  {\it stepwise increase} precisely at the factorizing point, due to the transition in the average level occupations. 
The negativities ${\cal N}_1$ and ${\cal N}_2$ (bottom left),  measuring the {\it pairwise} entanglement between first and second neighbors, exhibit instead a stepwise {\it decrease} at factorization, indicating multipartite entanglement effects of the parity projected states. They are also verified  to approach the {\it same side-limits} at factorization, confirming the independence from separation in its immediate vicinity, as  predicted  by the projected states \eqref{P}.      

In Fig.\ \ref{f4} we show the same quantities for a ring of  $N=6$ particles with the same  parameters, to view the trend for larger systems.  Their   behavior remains similar, 
with factorization located at the same point, where the four lowest levels with distinct parities cross (top left). However, the GS now exhibits in the range considered  two further parity transitions, at $v_{c2}\approx 1.52v_c$ and $v_{c3}\approx 1.74 v_c$, not related to factorization, where just two levels cross and the  GS parity changes from $(\sigma_2,\sigma_3)=(+,+)$ for $v<v_c$ to $(-,-)$ for $v_c<v<v_{c2}$, $(+,-)$ for $v_{c2}<v<v_{c3}$ and back to $(+,+)$ for $v>v_{c3}$.  

These transitions lead to further steps in the single site occupation numbers and entropy (right panels), though the larger step occurs again at the  factorizing transition. All three pair negativies ${\cal N}_i$ are  verified to reach the {\it same side-limits} at the factorizing point, a characteristic signature of uniform factorization,  exhibiting there a stepwise decrease. These patterns are not repeated at the other GS parity  transitions, where ${\cal N}_1$  increases but ${\cal N}_3$ decreases, vanishing for $v>v_{c3}$. 
Full range pairwise entanglement is thus centered at the factorizing point, where it becomes independent of separation. However,  the side-limits of ${\cal N}$ at factorization are smaller than for $N=4$, in 
agreement with monogamy and previous considerations. 

In Fig.\ \ref{f5} we show the  eigenvalues $p_i$ (entanglement spectrum) of the two-site density matrix $\rho_{pq}$  (left panel),  which determine the entanglement of the pair with the rest of the chain (just $4$ of them are nonnegligible). They also exhibit steps at the parity transitions, with the larger step again at the factorizing point. The ensuing mutual information   
\begin{equation}
    I_{pq}= S(\rho_p)+S(\rho_q)-S(\rho_{pq}) 
\end{equation} 
where $S(\rho_p)=S_p$  is the single  site entropy, is shown on the right panel for the first three neighbors. It is a measure of the total correlation between sites. It is seen that all three values merge at the side-limits of the factorizing point, confirming again that  in its vicinity correlations become  independent of separation.  Since it does not satisfy monogamy, its behavior is, however,  different from that of the  negativity, steadily increasing up to $v_{c2}$ and exhibiting at factorization a stepwise increase.

 \begin{figure}[htbp!]
\centerline{\scalebox{.6}{\includegraphics[trim={2 0 0 0},clip]
{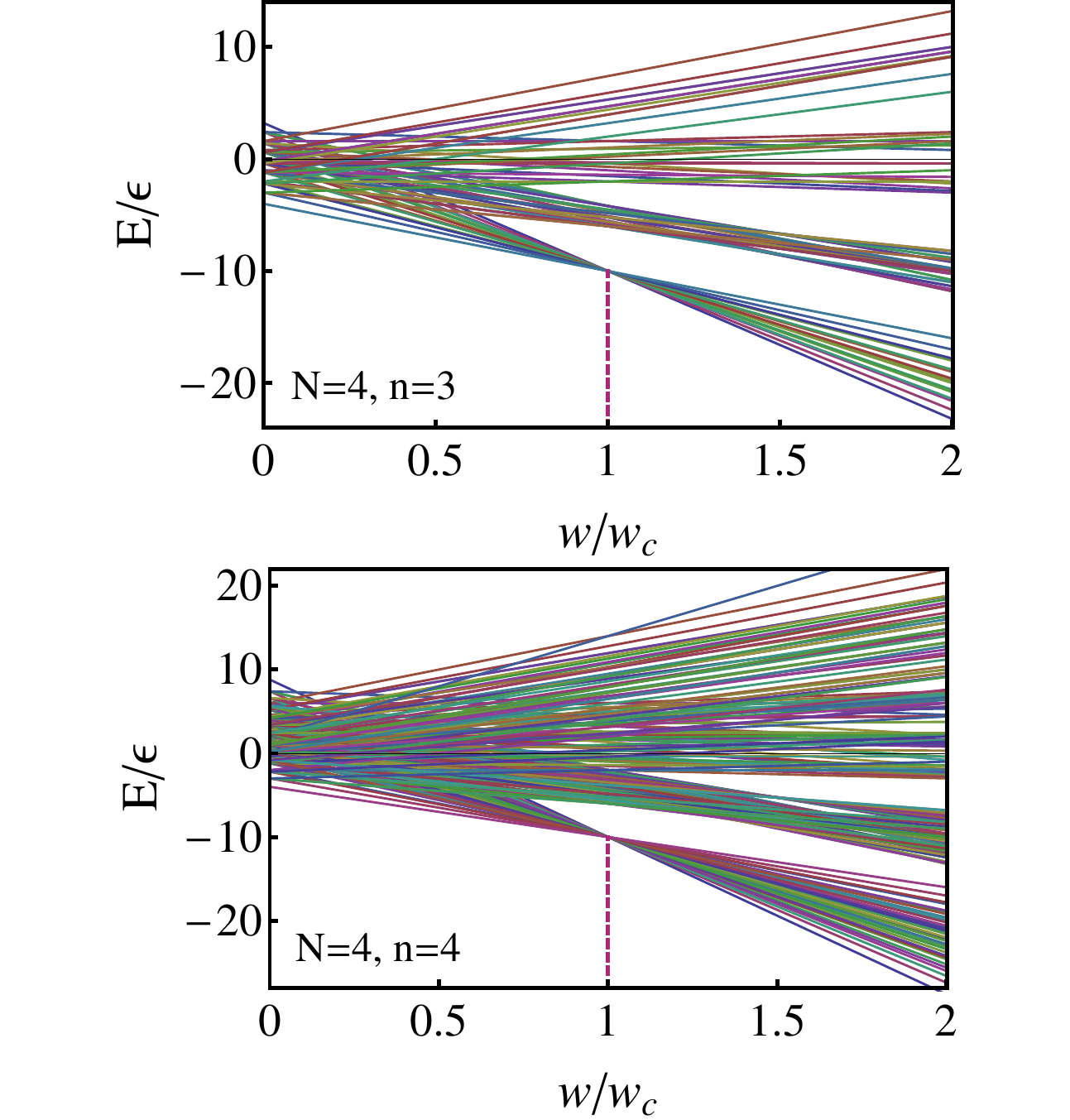}}}
\caption{The exact spectrum of  Hamiltonian \eqref{Hn3} for  $V_{ij}=0$ and first neighbor $W$ and $U$ couplings (see text), for  $N=4$ sites  and $n=3$ (top) and $4$ (bottom) levels at each site, as a function of  the scaled  coupling strength $w/w_c$.   Factorization arises at an exceptionally degenerate  point $w=w_c$ where  $15$ ($35$) levels  cross  for  $n=3$ ($4$), in agreement with Eq.\ \eqref{mw}. At this point {\it any} uniform factorized state is an exact GS.}
\label{f6}
\end{figure}

\begin{figure}[htbp!]
\centerline{\scalebox{.6}{\includegraphics[trim={2 0 0 0},clip]
{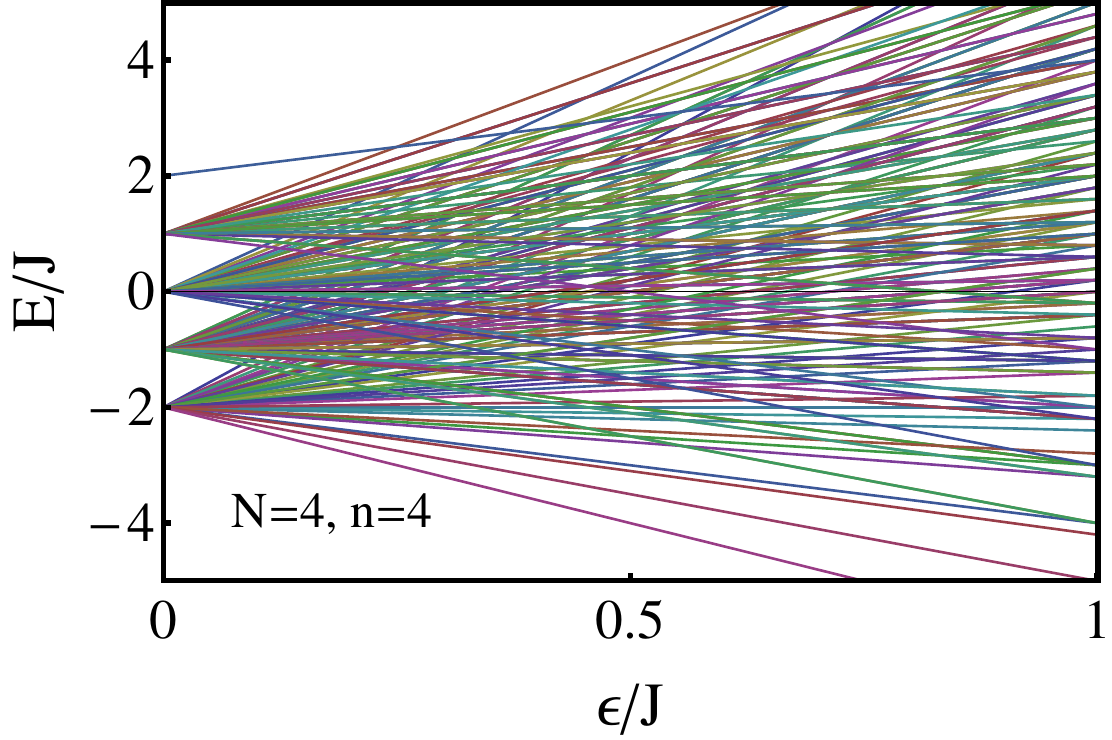}}}
\caption{The spectrum of  Hamiltonian \eqref{Hn3} for  $V_{ij}=0$ and  $U_{ii}=W_{ij}=J$ $\forall$ $i,j$, as a function of the single particle spacing $\epsilon/J$, for $N=n=4$ (see text). For $\epsilon\rightarrow 0$ the $SU(n)$ invariant Hamiltonian \eqref{HSun} is approached. In this limit any uniform factorized state is again an exact GS, with the GS degeneracy ($D=35$) given by the same Eq.\ \eqref{mw}.}
\label{f7}
\end{figure}

 Finally,  Figs.\ \ref{f6} and \ref{f7} show  the spectrum of $H$ in  the special $W$ case ($V_{ij}=0$) of sec.\ \ref{Wsec}, for $N=4$ particles and cyclic first neighbor couplings. In Fig.\ \ref{f6} 
 we consider $n=3$ (top) and $4$ (bottom) levels at each site, with uniform  spectrum   $\epsilon_1=-\epsilon$, $\epsilon_2=0$, $\epsilon_3=0.8\epsilon$ (and $\epsilon_4=2.2\epsilon$ for $n=4$), unequally spaced 
 in order to avoid extra  degeneracy away from factorization. We have set $U_{ij}=\delta_{ij}\frac{w}{w_c}(2\epsilon_i-E_2)$ and   $W_{ij}=\frac{w}{w_c} (\epsilon_i+\epsilon_j-E_2)$, 
 with $w_c=\epsilon$ and $E_2=-5\epsilon$,  such that  factorization takes place  at $w=w_c$  according to Eqs.\ \eqref{wii}--\eqref{jij2}, with GS energy  $\frac{N}{2}E_2=-\frac{5}{2}N\epsilon$,  independent of $n$.    

It is verified that all $\binom{N+n-1}{N}$  levels ($15$ for $n=3$ and 35 for $n=4$) forming the ``GS band'' cross at the factorization point $w=w_c$, where any uniform product state is confirmed to be  an exact GS.
The side-limits at $w=w_c$ of the crossing states   are  the symmetric states \eqref{Pns2} with definite  occupations in all $n$ levels, whose energies become all identical at this point,  with the GS changing at $w_c$  from $|\Psi_1\rangle$ (Eq.\ \eqref{psin}, all particles in the first level) to $|\Psi_n\rangle$ (all particles in the last level). No other multilevel crossing in higher excited states occurs at this point.

To complete the description, Fig.\ \ref{f7} depicts the spectrum for fixed couplings $W_{ij}=U_{ii}=J>0$ $\forall \, i,j$ and previous single site energies, as a function of the spacing $\epsilon$ for $n=4$ levels. At fixed $J$ factorization  is then  reached for $\epsilon\rightarrow 0$, where  $H$ becomes the $SU(n)$ invariant Hamiltonian  \eqref{HSun} and Eqs.\ \eqref{wii}--\eqref{jij2} are fulfilled, with $E_2=-J$ and  GS energy $-NJ/2$ $\forall$ $n\geq 2$. Again, all $35$ levels of the initial GS band  merge in this limit, where any uniform product state becomes an exact GS. 

However, in contrast with Fig.\ \ref{f6}, it is seen that the remaining higher energy levels also coalesce for  $\epsilon\rightarrow 0$ into  four levels, three of them highly degenerate (the highest level remains nondegenerate), 
due the high symmetry of $H$ for $\epsilon=0$.  Nevertheless, these higher energy eigenspaces contain no fully factorized states.  As can be seen from  \eqref{wii}--\eqref{jij2}, even if nonuniform product states were considered, no further fully separable eigenstate  is feasible for $\epsilon=0$, apart from those of the GS subspace. 
 
 For $N=4$ and $n\geq 4$,  the spectrum of Hamiltonian \eqref{HSun} with first neighbor couplings has just five distinct energies with uniform spacing: $E_i=-J(3-i)$ for $i=1,\ldots,5$. For $n=4$   the level degeneracies are $(35,110,60,50,1)$, the highest  level corresponding to the fully antisymmetric eigenstate. We remark, however, that while the same  factorized GS's  hold {\it also} in the presence of long range or nonuniform couplings, i.e.\ arbitrary $r_{pq}>0$,  with the same degeneracy 
 \eqref{mw} (and also  the {\it same} energy if  $r_p=\sum_{q\neq p} r_{pq}=1$ $\forall \, p$), the intermediate levels and degeneracies do depend on the coupling range and $r_{pq}$,  and are hence  not ``universal''. 
 Only the fully antisymmetric eigenstates, feasible for $n\geq N$, remain also unaltered, with an energy  which is just the opposite of that  of the fully symmetric  factorized eigenstates. 
 
\section{Conclusions\label{IV}}
We have analyzed the problem of GS factorization beyond the standard interacting spin system scenario. We have first derived general necessary and sufficient factorization conditions for Hamiltonians with two-site couplings, showing that they can be recast as pair eigenvalue equations. 
These conditions were then applied to interacting $N$-particle systems, where  
each constituent has access to $n$ local levels. 
For the $UVW$ class of  Hamiltonians \eqref{Hn3} they can be worked out explicitly, leading in the uniform case to the eigenvalue equation \eqref{eigg} for the squared local wave function and the constraint \eqref{jij} on the coupling strengths, valid for any number $n$ of levels. 
They are independent of size $N$ and coupling range, and generalize those for $XYZ$ spin systems, 
recovered for $n=2$. 
 The ensuing product state is shown to be a GS  when conditions  \eqref{ineq} are fulfilled, which are directly satisfied for vanishing $U_{ij}$. 
 
 The full rank factorized GS breaks all level number parities,  preserved by the Hamiltonian, therefore having a $2^{n-1}$ degeneracy (for $N\geq n-1$).  Factorization then arises at a special point where all $2^{n-1}$ definite parity levels of the GS band cross and become degenerate, signaling a fundamental GS  level parity  transition emerging for {\it any} size $N$ and range.  
 
 We have also examined the special $V=0$ case, where the Hamiltonian preserves the total occupation of each level.  Here the factorization conditions allowed us to identify  an  exceptional critical point, again emerging for any size and range, where {\it all levels} with definite occupations $N_i$ forming the GS band coalesce and become degenerate. This leads to a GS degeneracy which {\it increases with system size} ($D\propto  N^{n-1}$). At this point  {\it all  uniform product states}, including those breaking all occupation number symmetries, are exact  degenerate GSs, implying a full $SU(n)$ invariant GS subspace,  in a Hamiltonian which for $n\geq 3$ is not necessarily $SU(n)$ invariant. 
 
 Finally, we have  analyzed the entanglement properties in the immediate vicinity of factorization. For small systems, pairwise entanglement (as detected by the negativity) reaches there full range and becomes {\it independent} of separation,  thus constituting  an entanglement critical point.   Moreover, in such systems the parity transition occurring at the factorizing point entails finite discontinuities in most quantities (single site entanglement, negativity, level occupations, mutual information, etc.), whose magnitude can be analytically determined through projection of the factorized GS. On the other hand, for large systems pairwise entanglement will become vanishingly small  at factorization for any pair,   but  long range entanglement in its vicinity as well other effects (like bounded values of block entropies, Eq.\ \eqref{SMbnd}) will remain visible.  
 
In summary, in addition of providing nontrivial analytic exact GSs in strongly coupled systems which are not exactly solvable (which could be used as benchmarks for approximate numerical techniques),  symmetry-breaking factorization enables one to identify critical points in small samples with exceptional GS degeneracy and entanglement properties. 
Amidst increasing quantum control capabilities, present results  open the way to explore factorization in $SU(n)$ many-body physics and complex systems beyond the usual $SU(2)$ spin scenario. 

\acknowledgments 
Authors acknowledge support from CONICET (F.P. and N.C.) and CIC (R.R.) of Argentina.  Work supported by
CONICET PIP Grant No. 112201501-00732.
\appendix

\section{Special cases of Hamiltonian \eqref{Hn3}}
\label{ApA}
We consider here particular cases of   Hamiltonian \eqref{Hn3}. 
Fully connected fermionic  $U(n)$ nuclear models as those used in \cite{Mes.71,NPRC.85}, correspond to $r_{pq}=\frac{1}{N-1}$ $\forall$ $p\neq q$. In this case, for $U_{ij}=0$ and $\epsilon_i^p=\epsilon_i$ we can  rewrite \eqref{Hn3} as 
\begin{equation}
    H=\sum_{i=1}^n\epsilon_i G_{ii}-{\textstyle\frac{1}{2(N-1)}}
    \sum_{i\neq j} V_{ij}G_{ij}^2+W_{ij}(G_{ij}G_{ji}-G_{ii})
  \,,\label{Hnn}
\end{equation}
where $G_{ij}=\sum_{p=1}^\Omega   c^\dag_{pi}c_{pj}$ are collective operators satisfying the same $U(n)$ algebra as the  operators $g_{ij}=c^\dag_{pi}c_{pj}$: 
\[[G_{ij},G_{kl}]=\delta_{jk}G_{il}-\delta_{il}G_{kj}\,,\]
for both fermions and bosons. 
Eq.\ \eqref{Hnn} is a simplified schematic model for describing  collective  excitations. For $n=2$ and $\epsilon_i=(-1)^i\epsilon/2$ it becomes the   Lipkin Hamiltonian \cite{Lip,MDT.19}
\begin{equation*}
    H=\epsilon S_z -{\textstyle\frac{1}{2(N-1)}}[V(S_+^2+S_-^2)+W
    (S_+S_-+S_-S_+-N)]  
\end{equation*}
where $S_z=\frac{1}{2}(G_{22}-G_{11})$, 
$S_+=G_{21}=S_-^\dag$  are collective   spin operators satisfying the  $SU(2)$  algebra  ($[S_z,S_{\pm}]=\pm S_{\pm}$, $[S_-,S_+]=2S_z$)  and $V=V_{12}$, $W=W_{12}$.  These  models have been used to test several many-body techniques  \cite{Lip,NPRC.85,RR2.87,MDT.19,RS.80}, as the exact eigenstates   can be obtained by diagonalizing  $H$ in the irreducible representations of $U(n)$. 
For $n=2$ level number parity conservation reduces to the  $S_z$-parity symmetry $[H,P_z]=0$, 
 where $P_z=e^{-i\pi S_z}=P_2 e^{-i\pi N}$. 
 
 On the other hand, in the distinguishable formulation, the Hamiltonian \eqref{Hn3}  corresponds, for $g_{p}^{ij}=|i_p\rangle\langle j_p|$, to 
\begin{equation*}
    H=\sum_{i,p}\epsilon_i^p g_p^{ii}-\!\!
  \sum_{p<q,i,j}\!\!r_{pq} (U_{ij}g_p^{ii}g_q^{jj}+V_{ij}g_p^{ij}g_q^{ij}+W_{ij}g_p^{ij}g_q^{ji})\,.
   \end{equation*}
For $n=2$, $\epsilon^p_i=(-1)^ib^p/2$,   $V_{12}=(J_x-J_y)/2$, $W_{12}=(J_x+J_y)/2$  and 
$U_{11}=U_{22}=-U_{12}=J_z/2$, 
with $p=1,\ldots,N$, it becomes the Hamiltonian of $N$ spins $1/2$ interacting through anisotropic $XYZ$ couplings 
\cite{T.04,Baxter.71,RCM2.09,Z.21} of general range in a  nonuniform field $b^p$: 
\begin{equation} H=\sum_p b^p s_{pz}-\sum_{p\neq q}r_{pq}\!\!\sum_{\mu=x,y,z}  J_\mu s_{p\mu}s_{q\mu}\,,
\label{H2}\end{equation}
where $s_{pz}=\frac{g_p^{22}-g_p^{11}}{2}$,   $s_{px}=\frac{g_{p}^{21}+g_p^{12}}{2}$,
$s_{py}=\frac{g_{p}^{21}-g_p^{12}}{2i}$,
are spin operators satisfying the $SU(2)$ algebra. For $V_{12}=0$ we recover the $XXZ$ case where $J_x=J_y$ and $[H,S_z]=0$. 

Besides,  in the $n$-level case  the  operators $g_p^{ij}$ can always be expressed in terms of powers of spin-$s$ operators with $2s+1=n$. For instance, for $n=3$  all $g_p^{ij}$ can be written in terms of spin-$1$ operators $s_{pz}$ and $s_{p\pm}=s_{px}\pm is_{py}$ as  
\begin{eqnarray}
 g^{^{33}_{11}}_p &=&{\textstyle\frac{1}{2}}(s^2_{pz}\pm s_{pz})\,,\;\;
g^{22}_p={\textstyle\frac{1}{2}}s_p^2 -s^2_{pz}\,,\\
g^{21}_p&=&-{\textstyle\frac{1}{\sqrt{2}}}
s_{p+}s_{pz}\,,\;\;\;g^{32}_p={\textstyle\frac{1}{\sqrt{2}}}
s_{pz}s_{p+}\,,
\end{eqnarray}
with $g^{31}_p=\frac{1}{2}s_{p+}^2$,  $g_p^{ji}=(g_p^{ij})^\dag$ and  $s_p^2=s_{px}^2+s_{py}^2+s_{pz}^2=2\mathbbm{1}_p$. 
Thus, single site operators become 
in general quadratic in the local spin components $S_{p\mu}$. 

We now verify that for $n=2$, factorization conditions \eqref{eigg}--\eqref{jij} become those for the $XYZ$ Hamiltonian  
in a uniform field $b^p=b$ \eqref{H2}.  Eq.\ \eqref{eig1} leads for $n=2$ to 
\[E_2=-J_z/2-\sqrt{b^2+V_{12}^2}\,,\] 
for the lowest pair energy, with  \eqref{jij} implying 
$W_{12}=-E_2-U_{12}$. We then obtain 
\begin{equation*}
    |b|=\sqrt{(W_{12}-J_z)^2-V_{12}^2}=
    \sqrt{(J_y-J_z)(J_x-J_z)}\,,
\end{equation*}
 which is the known expression for the  factorizing field $b$ at  given couplings $J_\mu$ \cite{RCM.08,RCM2.09}  (valid for $J_z<J_y<J_x$, corresponding to $W_{12}>0$,  $V_{12}>0$).  Setting now $\bm{f}=(\cos
\frac{\theta}{2},\sin\frac{\theta}{2})$ for the local eigenvector, 
 Eq.\ \eqref{eig1}  leads to 
\begin{equation}
    \cos\theta={\textstyle\frac{b-J_z/2-E_2-V_{12}}{b-J_z/2-E_2+V_{12}}=\sqrt{\frac{J_y-J_z}{J_x-J_z}}}, 
\end{equation}
which coincides with the known expression for the spin orientation angle $\theta$ of the uniform product GS \cite{RCM.08}. 

In the $V=0$ case  of sec.\ \ref{Wsec},   factorization Eqs.\   \eqref{wii}--\eqref{jij2} imply  $U_{ii}=2\epsilon_i-E_2$ and  $W_{12}=-E_2-U_{12}\equiv J$ for $n=2$ and $\epsilon_2=-\epsilon_1$,  leading to a Heisenberg Hamiltonian 
\begin{equation}H=-\sum_{p\neq q}r_{pq}(J\bm{s}_p\cdot\bm{s}_q+C)\,,
\label{Heis2}\end{equation}
with $C=-\frac{1}{2}(E_2+\frac{1}{2}J)$.   
Both $E_2$ and  $U_{12}$ are free parameters.   
 It is verified that for $J>0$, 
 any uniform product state,  i.e.\ any state with all spins aligned in a fixed direction $\theta,\phi$ ($\bm{f}=(\cos\frac{\theta}{2},e^{i\phi}\sin\frac{\theta}{2})$) is an exact GS with pair energy $E_2$ ($\bm{s}_p\cdot\bm{s}_q|\psi,\psi\rangle=
\frac{1}{4}|\psi,\psi\rangle$) and total energy \eqref{EEE}. 
\section{Mean field approximation} 
 \label{ApB}
 We  show here that  the mean field (MF) approximation for the Hamiltonian \eqref{Hn3} (which corresponds to the Hartree-Fock (HF) scheme in the fermionic case)  
 can be solved analytically in the uniform attractive case, for any values of $n$, $N$ and the coupling range $r_{pq}\geq 0$.

 We look for the product state $|\Psi\rangle$ (or equivalently, the independent particle  state \eqref{psiuf})  which minimizes 
$\langle H\rangle=\langle\Psi|H|\Psi\rangle$ with $\epsilon_i^p=r_p\epsilon_i$ and nonegative couplings $U_{ij}$, $V_{ij}$, $W_{ij}$. 
As  $\langle c^\dag_{pi}c_{qj}\rangle= \delta_{pq}f^{p*}_if^p_j$ and 
$\langle c^\dag_{pi}c^\dag_{qj}c_{ql}c_{pk}\rangle=f_i^{p*}f_j^{q*}f_k^pf_l^q$ for $p\neq q$,  it is easily seen that in this case $\langle H\rangle$ can be  minimized by real uniform  coefficients  $f^p_i=f_i\in\mathbb{R}$. 
This leads, setting $r=\sum_p r_p=\sum_{p\neq q}r_{pq}$,    to 
 \begin{eqnarray} 
 \langle H\rangle
 & =&r(\sum_i \epsilon_i f_i^2-\frac{1}{2}\sum_{i,j} J_{ij}f_i^2f_j^2)\label{EHF2}\\
 &=&\frac{r}{2}\sum_{i,j} \tilde{M}_{ij}f_i^2 f_j^2,\;\; \tilde{M}_{ij}=\epsilon_i+\epsilon_j-J_{ij}\,,\quad\quad\label{A2}
  \end{eqnarray}
 where $J_{ij}=U_{ij}+V_{ij}+W_{ij}$ (and   $W_{ii}=V_{ii}=0$). Thus, MF depends here just on the sum of coupling strengths. 
 
 In order to obtain the MF solution, we may directly minimize \eqref{A2} with respect to the $f_i^2$, with the  constraint $\sum_i f_i^2=1$. After introducing a Lagrange multiplier $\lambda$, this leads to the equation  $\sum_j {\tilde M}_{ij}f_j^2=\lambda$  and hence to $f_i^2=\lambda\sum_{j}{\tilde M}^{-1}_{ij}$, i.e.\ $\bm{f}^2=\lambda \tilde{M}^{-1}\bm{v}$, with $\bm{v}=(1,\ldots,1)^T$.  
 Enforcing the constraint  leads to    $\lambda=1/(\bm{v}^T\tilde{M}^{-1}\bm{v})$ and 
 \begin{equation}
     \bm{f}^2=\tilde{M}^{-1}\bm{v}/(\bm{v}^T \tilde{M}^{-1}\bm{v})\,.\label{f2mf}
 \end{equation}
 The minimum MF energy becomes 
\begin{equation}
\langle H\rangle=\frac{r}{2}(\bm{f}^2)^T
{\tilde M}\bm{f}^2=\frac{r}{2}(\bm{v}^T{\tilde M}^{-1}\bm{v})^{-1}=\frac{r}{2}\lambda\,.\label{Emf}
\end{equation}
    Eqs.\ \eqref{f2mf}--\eqref{Emf} provide a closed expression for the full  parity breaking ($f_i\neq 0$ $\forall \, i$) MF state and energy. The sign of each $f_i$ remains free, in agreement with parity breaking, entailing a  $2^{n-1}$ degeneracy of the MF state. 
    
  The exact factorized GS determined by Eqs.\ \eqref{eigg}--\eqref{jij} is one of these solutions: at factorization, \eqref{jij} implies $J_{ij}=\epsilon_i+\epsilon_j-E_2+V_{ij}$ for $i\neq j$ and hence 
\begin{eqnarray}
   {\tilde M}_{ij}&=&(2\epsilon_i-U_{ii})\delta_{ij}
  -(1-\delta_{ij})(V_{ij}-E_2)\nonumber
  \\&=&M_{ij}+E_2(1-\delta_{ij})\,,  \label{A5}\end{eqnarray}
   with $M$ the matrix in   \eqref{eig20}. Eqs.\   \eqref{f2mf}--\eqref{A5} imply   
  Eq.\ \eqref{eigg}, with    $E_2=(\bm{v}^T{\tilde M}^{-1}\bm{v})^{-1}=\lambda$  the MF pair energy.  

 The restriction $f_i^2> 0$ $\forall i$ 
 implies, however,  a limit on the validity of solution \eqref{f2mf}. The border is obtained from the condition $f_i=0$ for some $i$ (normally the highest energy level). Beyond this border we should set $f_i=0$,  obtaining a new MF solution with $n-1$ occupied levels, given by \eqref{f2mf} with $\tilde{M}$, $\bm{v}$ restricted to the occupied levels. This solution is valid until one of the new coefficients $f_i^2$ vanishes. For decreasing coupling strengths, this is to be  repeated  until the trivial  solution $f_i=\delta_{i1}$ (valid for sufficiently small $J_{ij}$) is reached. 
 
Therefore, as $J_{ij}$ increases from $0$, a  series of $n-1$ MF transitions normally arise, associated with the onset of occupation of the $i_{\rm th}$ level.
For instance, for $U_{ii}=0$ and 
$J_{ij}=J(1-\delta_{ij})$, $J>0$, Eq.\ \eqref{f2mf} leads to 
\begin{equation}
    f_i^2=1/n-\tilde{\epsilon}_i/J\,,\;\;i=1,\ldots,n\,,\label{fi2}
\end{equation}
where $\tilde{\epsilon}_i=\epsilon_i-\frac{1}{n}\sum_{j=1}^n\epsilon_j$ is the centered spectrum ($\sum_{i=1}^n\tilde{\epsilon}_i=0$).  Eq.\ \eqref{fi2} 
holds insofar $f_i^2\geq 0$ $\forall$ $i$, i.e.\ 
\begin{equation}
    J\geq J^c_n=n\tilde{\epsilon_n}\,\label{Jcn}\end{equation}
where $n\tilde{\epsilon}_n=\sum_{j=1}^{n-1}\epsilon_n-\epsilon_{j}$ is the sum of energy differences with all lower levels. 
Repeating the procedure for a solution with just the first $m$ levels  occupied, 
the same expressions \eqref{fi2}--\eqref{Jcn}  are obtained  
  with $n\rightarrow m$. 
  
  \section{Splitting of energy levels at the border of  factorization\label{ApC}}
Let us assume that $H=H_f+\delta H$, where 
$H_f=H_0+V_{\rm int}$ is the Hamiltonian having the factorized GS and  
\begin{equation}
\delta H_0=\sum_{i}\delta\epsilon_{i}\sum_{p}
c^\dag_{pi}c_{pi}
\end{equation}
a small perturbation of the single particle term. For instance, a perturbation $\delta V_{int}=\gamma V_{int}$ leads to $\delta H=\gamma H_f-\gamma H_0$, implying $\delta\epsilon_i=-\gamma\epsilon_i$ plus a constant energy shift $\delta E=\gamma E_f$. 
 At first order in $\delta \epsilon_i$, the remaining correction on the definite parity energy levels is 
\begin{equation} \delta E_{\sigma_2,\ldots,\sigma_n}={\textstyle\sum_{i}}\delta\epsilon_i \langle N_i\rangle_{\sigma_2,\ldots,\sigma_n}\label{C2}\,,
\end{equation}
where $N_i=\sum_pc^\dag_{pi}c_{pi}$ and the average is taken on the parity projected states \eqref{P}. For    $n=3$, $\langle N_i\rangle_{\sigma_2,\sigma_3}/N$ is given in Eq.\ \eqref{fp}. We then obtain, setting $u_j=1-2|f_j|^2$, 
\begin{eqnarray}
{\textstyle\frac{\delta E_{\sigma_{2}\sigma_{3}}}{N}}&=&
{\textstyle\frac{\sum_{i}\delta\epsilon_{i}|f_i|^2\left[1+\sum_{j}\sigma_{j}\left(-1\right)^{\delta_{ji}}u_j^{N-1}\right]}{1+\sum_{j}\sigma_{j}u_j^N}}\nonumber\\
&\approx&{\textstyle\sum_{i}\delta\epsilon_{i}|f_i|^2[1+\sum_j\sigma_j((-1)^{\delta_{ji}}+2|f_j^2|-1)u_j^{N-1}]}\nonumber
\end{eqnarray}
where $\sigma_{1}\sigma_{2}\sigma_{3}=\left(-1\right)^{N}$ and  last expression holds for sufficiently large $N$. 
For $\delta \epsilon_3=-\delta\epsilon_1=\delta\epsilon$ and $\delta\epsilon_2=0$, 
this leads to  $\delta E_{++}<\delta E_{-+}<\delta E_{+-}<\delta E_{--}$ for $\delta \epsilon>0$.  This is the case of Fig.\ \ref{f2}, where 
$\delta\epsilon=(1-\frac{v}{v_c})\epsilon>0$ ($<0$) on the left (right) side of the factorization point $v=v_c$. 
In the $V=0$ case, $\langle N_i\rangle=n_i$ is just the occupation of level $i$ in the projected states \eqref{Pns}--\eqref{Pns2}, and  \eqref{C2} becomes exact. 
%

\end{document}